\documentclass[%
 reprint,
 superscriptaddress,
 nofootinbib,
 amsmath,amssymb,
 aps,
 prx,
floatfix,
]{revtex4-2}
\pdfoutput=1
\usepackage[utf8]{inputenc}
\usepackage[english]{babel}
\usepackage[T1]{fontenc}
\usepackage{comment}
\usepackage{times}

\usepackage{hyperref}
\usepackage{amsthm}
\usepackage{amsthm}
\usepackage{amsfonts}
\usepackage{braket}
\usepackage{enumerate}
\usepackage{graphicx}
\usepackage{bbm}
\usepackage{stmaryrd}
\usepackage{nicefrac}
\usepackage[noend,linesnumbered, ruled, vlined]{algorithm2e}

\usepackage[dvipsnames]{xcolor}
\hypersetup{
    colorlinks=true,
    linkcolor={teal}, 
    citecolor=Maroon,     %
    filecolor=blue,     %
    urlcolor=Maroon,      %
}

\pdfoutput=1

\input{preamble.sty}
\newcommand{\secref}[1]{Section~\hyperref[#1]{\ref*{#1}}}
\newcommand{\appref}[1]{Appendix~\hyperref[#1]{\ref*{#1}}}
\newcommand{\tabref}[1]{Table~\hyperref[#1]{\ref*{#1}}}
\newcommand{\figref}[1]{Figure~\hyperref[#1]{\ref*{#1}}}
\newcommand{\sfigref}[2]{Figure~\hyperref[#1]{\ref*{#1}(#2)}}
\newcommand{\boxref}[1]{Box~\hyperref[#1]{\ref*{#1}}}
\newcommand{\algref}[1]{Algorithm~\hyperref[#1]{\ref*{#1}}}
\newcommand{\thmref}[1]{\hyperref[#1]{Theorem~\ref*{#1}}}

\newcommand{\vect}[1]{\boldsymbol{#1}}

\begin{document}

\title{Quantum low-density lattice codes}

\author{Timo Hillmann}
\email{timo.hillmann@rwth-aachen.de}
\affiliation{School of Physics, University of Sydney, Sydney, NSW 2006, Australia}
\affiliation{Department of Microtechnology and Nanoscience (MC2), Chalmers University of Technology, SE-412 96 Gothenburg, Sweden}

\author{Jens Eisert}
\email{jense@zedat.fu-berlin.de}
\affiliation{Dahlem Center for Complex Quantum Systems, Freie Universit{\"a}t Berlin, 14195 Berlin, Germany}
\affiliation{Helmholtz-Zentrum Berlin f{\"u}r Materialien und Energie, 14109 Berlin, Germany}

\author{Francesco Arzani}
\email{francesco.arzani@inria.fr}
\affiliation{QAT team, DIENS, \'Ecole Normale Sup\'erieure, PSL University, CNRS, INRIA, 45 rue d'Ulm, Paris 75005, France}

\begin{abstract}
Gottesman–Kitaev–Preskill (GKP) codes provide a 
family of promising schemes for encoding discrete quantum information (qudits) into infinite-dimensional bosonic modes based on mathematical lattices. 
While such codes, when concatenated with discrete-variable codes, are relatively well studied, the construction and decoding of native GKP codes has largely remained open due to the computationally hard problems encountered. 
To address this challenge, we advocate a strategy of co-designing the decoder and the quantum error-correcting code itself by constructing lattices for which decoding is feasible: The requirement of efficient decoding effectively determines the quantum error-correcting code. 
This construction is built on classical low-density lattice codes (LDLCs), a lattice analogue of low-density parity-check codes, here lifted to families of GKP codes. 
Concretely, we introduce quantum versions of classical, randomly constructed LDLCs. 
We show that after suitable dimensionality reduction these codes have code properties comparable to or better than concatenated GKP-surface codes of equal number of modes. However, the GKP-LDLCs constructed here do not have a strictly sparse parity check matrix, which motivates our study of the performance of natively analog message-passing decoders originally developed for LDLCs when applied to concatenated GKP-LDPC codes. 
We show that the fully analog, linear-time decoder achieves performances close to state-of-the-art hybrid qubit-analog decoders. 
To facilitate future research on the structure and performance of general GKP codes, the relevant source code will be released in open-source Julia packages \textsc{LatticeDecoder.jl} and \textsc{SymplecticGKP.jl}.
\end{abstract}

\maketitle
\section{Introduction}\emph{Gottesman-Kitaev-Preskill} (GKP) codes \cite{GKP} are a bosonic version of stabilizer codes that represent
highly promising candidates for hardware-efficient fault-tolerant quantum computing~\cite{RevModPhys.87.307,Roads,QECBasic,eisert2025mindgapsfraughtroad}.
They encode finite-dimensional quantum systems into the infinite-dimensional Hilbert spaces of bosonic modes, where code states are superpositions of peaked wave functions whose translation symmetries in phase space correspond to mathematical lattices~\cite{Conrad_2022}. 
Such continuous-variable quantum error correcting codes are instrumental in several promising platforms for quantum computing, including  instances of superconducting~\cite{GKPSuperconducting,PhysRevA.101.053840}
architectures, systems of trapped ions~\cite{GKPIons}
or optical platforms~\cite{PhysRevLett.128.170503, GKPBlueprint, GKPLight, aghaee_rad_scaling_2025}.

A widespread construction to generate GKP codes relies on encoding a single qubit in each of many harmonic oscillators and then concatenating it with a qubit-level error-correcting code~\cite{PhysRevX.8.021054, PRXQuantum.3.010315, QLDPCGKP,PRXQuantum.4.020342,ReviewGKP,PRXQuantum.5.020349}. 
This approach leverages, on the one hand, established tools for qubit stabilizer codes, but at the same time leaves open the question whether more efficient schemes might exist that are \textit{not} based on concatenation. 
One reason to think that this might be the case is that the most relevant noise models for qubits are local, affecting only a few qubits at a time. 
On the other hand, most continuous-variable channels affect \textit{all} data carriers at the same time, but only cause small perturbations. 
The error digitization that converts small perturbations to low-weight errors in the qubit case does not happen in bosonic systems, as measurements give continuous outcomes, so the probability of projecting a single oscillator on exactly no error is zero. 
The effectiveness of discrete-variable techniques observed in numerical experiments~\cite{toricGKP, Noh_2019,hanggli_enhanced_2020-1,PRXQuantum.3.010315} is thus not theoretically well understood, which motivates the exploration of native continuous-variable code constructions.
Further motivation to consider non-concatenated constructions comes from decoders, algorithms that provide the correction given the syndrome.
The qubit structure in concatenated GKP-stabilizer codes is often exploited for decoding as well, but it has been shown that adapting these tools to take advantage of the continuous nature of the underlying bosonic observables can sensibly improve the logical error 
rate~\cite{PhysRevX.8.021054, PRXQuantum.3.010315, QLDPCGKP,PRXQuantum.4.020342,ReviewGKP,PRXQuantum.5.020349}.

This raises the question of whether a native analog strategy might perform better than hybrid solutions.
However, fully analog decoding in native GKP codes is challenging due to a severe bottleneck: standard decoding methods amount to solving computationally hard lattice problems. 
Indeed, a handful of previous works have explored this possibility that eventually involves solving some form of the \emph{closest-vector problem}  (CVP)~\cite{HarringtonRates,Conrad2024goodgottesmankitaev,Lin_2023}.
This is an NP-hard problem in worst-case complexity, 
as an exact problem, but it is also hard to approximate~\cite{aggarwal2015solvingclosestvectorproblem,
goldreich_approximating_1999,arora_hardness_1993}.
In fact, many lattice-based post-quantum cryptographic schemes rely on the hardness of the CVP or closely related problems. 
This complexity renders naive approaches to the decoding problem for GKP codes infeasible.

To overcome this bottleneck, we advocate a \emph{co-design of the code and the decoder}. 
Rather than designing a code first and subsequently searching for an efficient decoder, we reverse this perspective: the requirement of efficient decoding largely dictates what constitutes a useful quantum code in the first place.
To do this, we translate to the bosonic setting techniques inspired by classical \textit{low-density lattice codes}~\cite{ldlc}, lattice codes analogous to \emph{low-density parity check} (LDPC) codes. 
Classical LDLCs perform very close to the capacity of the additive noise random Gaussian channel with a linear-time iterative \emph{message-passing} (MP) decoder. 
The latter produces an approximate solution to CVP leveraging a sparsity condition in the lattice's parity check matrix. 
Put differently, the instances provided by such lattices are not hard instances of the general CVP.

First, we construct \emph{quantum} versions of \emph{LDLCs} (qLDLCs) and compute their code distance. 
The simplest construction we examine leads to typically very high logical dimension, but a proportionally small distance, resulting in poor error correction performance. 
We then detail a procedure to systematically construct codes with smaller logical dimension and higher distances. 
The price to be paid is the introduction of a few (two) non-sparse checks, making the corresponding codes less amenable to MP decoding, since the dense checks introduce short cycles in the decoding graph. Nevertheless, we show that they have distance comparable to rotated surface codes with the same number of modes, and that they lead to lower error rates than \emph{larger} surface codes when decoded through brute force closest-vector decoding. 

Secondly, we perform numerical experiments applying a version of the classical MP decoder to GKP codes concatenated with LDPC stabilizer (qubit) codes. 
While being different from LDLCs, these codes share the property of featuring a sparse parity check matrix.
The upshot is that the decoder is fully analog and natively incorporates the continuous information coming from GKP syndrome measurements. 
This contrasts with the necessity of \textit{ad hoc} modifications in other approaches to decode concatenated GKP-stabilizer codes. 

Our results show that a co-design of codes and decoders is needed to overcome obstructions in the decoding of native GKP codes based on lattices.
More broadly, it provides further evidence that the co-design of quantum error correcting codes on the one hand and architectures, schemes, and decoders, on the other hand, are a fruitful way towards achieving a fault-tolerant application-scale quantum computer~\cite{eisert2025mindgapsfraughtroad}. 

\section{Preliminaries and notation}

\subsection{Classical low-density lattice codes}

\subsubsection{Definition}

Low-density lattice codes, first introduced and analysed by Sommer \emph{et al.}~in Ref.~\cite{ldlc}, are a class of classical lattice codes characterized by sparse parity-check matrices. 
Their interest lies in the fact that they have asymptotically good code parameters and admit a decoder that runs in linear time with the lattice dimension, achieving error probabilities within a fraction of a dB of the capacity of the (classical) \emph{additive white Gaussian noise} (AWGN) channel.

\begin{mydef}[LDLC \cite{ldlc}]
A (classical) low-density lattice code is an $n$-dimensional lattice 
\begin{align*}
    \Lambda = \{G \vect{z}: \vect{z} \in \mathbb{Z}^n \},
\end{align*}
where the generator matrix $G \in \mathbb{R}^{n \times n}$ satisfies $\lvert \det{G} \rvert = 1$, and whose parity-check matrix $H = G^{-1}$ is sparse.
\end{mydef}

\noindent In other words, the Euclidean dual lattice of an LDLC admits a sparse generator. 
In this context, sparsity only makes sense in the asymptotic setting, where for a family of codes of dimension $n\to \infty$ the degree of each row and column of $H$ is bounded by some constant.  

It is possible to construct LDLCs based on randomized approaches. 
We follow the approach presented by Sommer \emph{et al.}~\cite{ldlc} to construct a regular, so-called, magic square LDLC, but point out that more elaborate construction techniques exist, see, e.g., Ref.~\cite{li_construction_2015}.
Conceptually, the randomized construction starts from a random permutation matrix of size $d \times n$, where $d$ specifies the row (and column) degree of the parity-check matrix $H$ to be constructed.
The entries of the permutation matrix are iteratively swapped until the corresponding decoding graph attains a girth of at least six.
This defines the binary support of the parity-check matrix $H$. 
In a second stage, this support is instantiated as a real-valued sparse matrix by assigning nonzero entries from a small, predefined set of coefficients $\{h_1, h_2,\dots, h_d\}$ of size $d$ with $h_1 \geq h_2 \geq \dots \geq h_d > 0$ with random signs.
The detailed description of the algorithm is provided in Appendix~\ref{sec:randLDLC}.

\subsubsection{Decoding \label{sssec:ldlc_decoding}}

In the classical setting, the code consists of the direct lattice $\mathcal{C} = \lrc{\aa^T G: \aa \in \mathbb{Z}^n}$. 
The encoded message is a lattice point $\xx \in \mathcal{C}$, and following the AWGN channel, the received vector is 
\begin{equation}
\yy = \xx + \ee, \quad \ee \sim \mathcal{G}_{\sigma,0},
\end{equation} 
where $\mathcal{G}{\sigma,0}$ denotes the zero-mean Gaussian distribution with covariance matrix $\sigma^2 \mathbf{I}$.
The classical maximum-likelihood (ML) decoder then outputs a solution $\bar{\xx} = \argmin_{\xx \in \mathcal{C}} \|\yy-\xx\|$ to the \emph{closest-vector problem} (CVP).
In practice, a finite set of lattice points is used to encode information, determined by the so-called shaping region.
However, it is common to benchmark the codes assuming that the shaping region is the whole space $\mathbb{R}^n$, that is, any lattice point might be sent. 

Solving the \emph{closest vector problem} (CVP) is NP-hard in worst-case complexity in general~\cite{MicciancioGoldwasser:2002}, and remains hard even to approximate within small quasi-constant~\cite{goldreich_approximating_1999, dinur_approximating_2003}
or even constant~\cite{arora_hardness_1993}
factors. 
Unsurprisingly, in light of the importance of the problem, there is a large body of literature that deals with heuristics of solving the CVP.
There are polynomial-time algorithms for large approximation factors such as
Babai’s nearest-plane algorithm~\cite{babai_lovasz_1986}.
Using a stronger reduction, one can improve on Babai’s output in sieving-based algorithms, often via enumeration in a reduced basis~\cite{Ajtai}. However, the worst-case runtime of the best known algorithms remains exponential in the lattice dimension. 

Instead of employing a CVP solver directly, the premise of LDLCs is to deliberately choose lattices for which a message-passing algorithm is effective. 
While this does not imply that the output is the closest vector to the received message, it will typically be a good approximation.

Specifically, the iterative decoder presented below exploits the sparsity of the parity-check matrix $H = G^{-1}$ to produce a candidate solution at a computational cost that scales linearly with the lattice dimension.
Each valid lattice point $\vect{x} = G\vect{a}$ satisfies by construction
\begin{align}
H\xx = HG\aa = \aa\in\mathbb{Z}^n\quad \forall \vect{x}\in\mathcal{C}. 
\end{align}
Therefore, the decoder can equivalently seek an integer vector $\vect{a}$ that is consistent with the noisy observation, also known as the received vector, $\vect{y}$.

The decoder first computes an approximation to the marginal probability distributions $p_j(\tilde{x}_j)$ for the components of the candidate codeword $\tilde{x}_j$, then these functions are maximized to obtain
\begin{align}
\tilde{\bar{\xx}} = \argmax_{\tilde{\xx}}\prod_jp_j(\tilde{x}_j).    
\end{align}
The resulting vector is not necessarily a lattice point, so a hard decision is made by rounding
\begin{align}
\label{eq:hard_decision}
\hat{\aa} = \lfloor H\tilde{\bar{\xx}} \rceil, \quad \hat{\xx} = G\hat{\aa}.
\end{align}

The iterative decoder operates on a bipartite graph, known as the decoding graph, which is specified by the sparse parity-check matrix $H$.
For each column of $H$, the decoding graph contains a variable node $v_k$, ultimately corresponding to a component $x_k$ of the codeword $\xx$.
Similarly, for each row of $H$, the decoding graph contains a check node $c_j$ that enforces the linear constraint $(H \xx)_j \in \mathbb{Z}$.
A variable node $v_k$ and a check node $c_j$ are connected if $H_{j,k} \neq 0$.

Using the definitions of check and variable nodes, the message-passing algorithm contains the following steps.%

\begin{enumerate}
    \item \textbf{Initialization.} Each variable node $v_k$ sends to all of its neighboring check nodes the initial message
    \[
    f_k^{(0)}(x) = \frac{1}{\sqrt{2\pi\sigma_k^2}} \exp\Big[-\frac{(y_k - x)^2}{2\sigma_k^2}\Big],
    \]
    where $y_k$ is the $k$-th component of the received vector, and $\sigma_k^2$ is the AWGN variance for the $k$-th component. These represent the channel prior distributions.

    \item \textbf{Check-to-variable messages.} Each check node $c_j$ sends a message to each neighboring variable node $v_{i_k}$. 
    The message $g_{j\to i_k}$ is computed in three sub-steps:
    \begin{description}[font=\normalfont\itshape]
        \item[Convolution:] Compute the convolution of the local messages of the $t$-th iteration
        \[
            p_{j\to i_k}(x) = \bigstar_{i \in \mathcal{N}(j) \setminus \{i_k\}} f_i^{(t)}\Big(\frac{x}{H_{j , i_k}}\Big), \label{eq:ldlc_alg_conv}
        \]
        where $\mathcal{N}(j)$ is the set of variable nodes connected to $c_j$.
        
        \item[Stretch:] Apply a linear transformation
        \[
            p'_{j\to i_k}(x) = p_{j\to i_k}(-H_{j , i_k} x).
        \]
        
        \item[Periodic extension:] Enforce lattice periodicity
        \[
            g_{j\to i_k}(x) = \sum_{l \in \mathbb{Z}} p'_{j\to i_k}\Big(x - \frac{l}{H_{j , i_k }}\Big).
        \]
    \end{description}
    
    \item \textbf{Variable-to-check messages.} 
    After receiving all incoming messages $g_{i_j \to k}(x)$, variable node $v_k$ computes the outgoing message to each check $c_{i_j}$ as
    \begin{description}[font=\normalfont\itshape]
        \item[Product:] 
        \[
            q_{k\to i_j}(x) = f_k^{(0)}(x) \prod_{l \in \mathcal{N}(k)\setminus \{i_j\}} g_{i_l \to k}^{(t-1)}(x).
        \]
        \item[Normalization:] 
        \[
            f_{k\to i_j}^{(t)}(x) = \frac{q_{k\to i_j}(x)}{\int_{\mathbb{R}} q_{k\to i_j}(z) \, dz}.
        \]
    \end{description}

    \item \textbf{Iteration.}
    Repeat steps 2 and 3 for a specified number of iterations.

    \item \textbf{Final marginal computation.} Each variable node computes its final estimate
    \[
        f_k^F(x) = f_k^{(0)}(x) \prod_{l \in \mathcal{N}(k)} g_{i_l \to k} (x).
    \]
\end{enumerate}

The message passing algorithm can be derived as in the discrete-variable instance, following Gallager's derivation~\cite{gallager_low_1960} by extending to the case of continuous marginal (prior) distributions.
Similarly, the decoder has linear complexity in the number of variable nodes.
Intuitively, the check-to-variable step corresponds to calculating the \emph{probability density function} (PDF)  for the variable $x_{i_k}$ from the neighboring variables $x_{i_j}, j \neq k$, under the constraint $\sum_{l = 1, l \neq k}^{d} H_{j, i_l} x_{i_l} \in \mathbb{Z}$ and the assumption that the $x_{i_l}$ are independent.
From the constraint follows that 
\begin{equation}
x_{i_k} = (z - \sum_{l = 1, l \neq k}^{d} H_{j ,i_l} x_{i_j}) / H_{j , i_k}, z\in \mathbb{Z}, 
\end{equation}
i.e., $x_{i_k}$ is a sum of (scaled) random variables.
The PDF of a sum of random variables is given by the convolution of the corresponding PDFs.
This is implemented by the equation just above,
assuming $z = 0$ and the stretching step.
In the last sub-step, the result is extended to 
arbitrary $z \in \mathbb{Z}$.
In contrast, the variable-node update corresponds to a Bayesian combination of independent sources of information about the same variable $x_{i_j}$.
Therefore, the outgoing message is simply the product of the incoming PDFs and the channel PDF.

Note that the convolution of two Gaussians with means $m_1$ and $m_2$ and variances $\Delta_1$ and $\Delta_2$ is another Gaussian with mean $m = m_1 + m_2$ and variance $\Delta = \Delta_1 + \Delta_2$~\cite{papoulis_probability_2002}.
Similarly, the product of two Gaussians is a scaled Gaussian with variance $\Delta = (\Delta_1^{-1} + \Delta_2^{-1})^{-1}$, mean $m = \Delta (m_1 \Delta_1^{-1} + m_2 \Delta_2^{-1})$, and amplitude 
\begin{equation}
A = \exp\left(-\frac{1}{2}\frac{(m_1 - m_2)^2}{\Delta_1 + \Delta_2}\right) / \sqrt{2\pi (\Delta_1 + \Delta_2)}.
\end{equation}
Thus, throughout the algorithm, all messages can be described as mixtures of Gaussians of the form $M_{\Delta}(x) = \sum_{j=1}^{\infty} A_j G(x, m_j, \Delta)$, and within a mixture all Gaussians have the same variance~\cite{ldlc}.
Hence, it is possible to analyze the convergence of the algorithm solely through the first and second moments, namely the means $m_j$ and variances $\Delta_j$, of the Gaussians and their amplitudes $A_j$.
Indeed, this statement is independent of the specific structure of the LDLC, and
Ref.~\cite{ldlc} merely tracks the first and second moments in the convergence proof.

A more precise statement about the dynamics of the decoder can be made by restricting to the case of a magic square LDLC with generating sequence $h_1 \geq h_2 \geq \dots \geq h_d > 0$.
In this case,  firstly, it can be shown that variable node messages sent during the same iteration along edges with the same absolute weight have the same variances~\cite[Lemma 3]{ldlc}.
Hence, we may denote the variance of the messages sent at iteration $t$ along edges with weight $\pm h_l$ by $\Delta_l^{(t)}$.
Secondly, it is seen that those $\Delta_1^{(t)},\dots, \Delta_d^{(t)}$ satisfy a simple recursion relation
\begin{equation}
\frac{1}{\Delta_i^{(t+1)}}
= \frac{1}{\sigma^2}
+
\sum_{m=1, m\leq j}^{d}
\frac{h_m^2}{
\sum_{j=1,j\leq m}^d
h_j^2 
\Delta_j^{(t)}
}\,.
\end{equation}
Using these insights, and in the same notation, the following theorem regarding the convergence of the variances can be derived.

\begin{theorem}[{\cite[Theorem 1]{ldlc}}]
    \label{thm:variance_convergence}
    For a magic square LDLC with generating sequence $h_1 \geq h_2 \geq \dots \geq h_d$, define $\alpha %
    \coloneqq
    \sum_{i=2}^{d} h_i^2 / h_1^2$ and assume $\alpha < 1$. 
    Then the steady-state of the variances $\Delta_i^{(\infty)}$ is given by
    \begin{align}
        \Delta_i^{(\infty)} := \lim_{t \to \infty} \Delta_i^{(t)} = \begin{cases}
            (1 - \alpha) \sigma^2, &i = 1, \\
            0, & i = 2, \dots, d,
        \end{cases}
    \end{align}
    where $\sigma^2$ is the channel variance.
    Furthermore, the asymptotic convergence rate of all variances is exponential, that is, 
    \begin{align}
        0 < \lim_{t \to \infty} \left\lvert \frac{\Delta_i^{(t)} - \Delta_i^{(t)}}{\alpha^t}  \right\rvert < \infty,
    \end{align}
    for $i = [d]$.
\end{theorem}

We note that in the case of $\alpha \geq 1$ the variances can still converge, although with an algebraic convergence rate $o(1/t)$~\cite{ldlc}.
Interestingly though, for the case that $h_1 = h_2 = \dots = h_d$, $\Delta_i^{(t)} = {\sigma^2}/{(t + 1)}$, all variances approach zero~\cite{ldlc}. 
For the following analysis, we will assume $\alpha < 1$.

\thmref{thm:variance_convergence} shows that during each iteration, variable nodes send $d-1$ messages with variance asymptotically zero and a single message with variance asymptotically constant sent along the edge with weight $h_1$.
On the other hand, check nodes generate $d-1$ messages with variances asymptotically constant and a single message with variance asymptotically zero.
Intuitively, the difference originates because in the convolution of two Gaussians, the new variance is obtained from the (ordinary) arithmetic sum of the input variances, in contrast to the product of two Gaussians, where the new variance is the harmonic sum of the input variances.

The separation between messages generated at variable nodes with variance asymptotically zero, called \emph{narrow} messages, and messages with variance asymptotically constant, called \emph{wide} messages, helps analyse the convergence of the mean values.
In particular, Sommer \emph{et al.} show that asymptotically, the mean values of the $d-1$ narrow messages become equal~\cite[Lemma 5]{ldlc}.
In particular, one can show within this analysis that a necessary condition for the narrow messages to converge to a lattice point requires that the spectral radius of $\tilde{H}$ fulfills $\rho(\tilde{H}) < 1$~\cite[Theorem 2]{ldlc}.
Here, the matrix $\tilde{H}$ is obtained from the parity-check matrix $H$ by permuting its rows such that the $\pm h_1$ entries are placed on the diagonal, normalizing each row by its diagonal element, and subsequently nullifying the diagonal.
Additionally, convergence of the wide messages to a lattice point requires that the spectral radius of $F$ fulfills $\rho(F) < 1$~\cite[Theorem 3]{ldlc}, where the matrix $F \in \mathbb{R}^{n \times n}$ is defined as
\begin{equation}
    F_{k,l} =
    \begin{cases}
        \frac{H_{r,k}}{H_{r,l}}, &\begin{array}{l}
             k \neq l \text{ and there exists a row } r \text{ of } H, \\
             \text{such that } |H_{r,l}| = h_1 \text{ and } H_{r,k} \neq 0,
        \end{array} \\
        0, & \text{otherwise}.
    \end{cases}
\end{equation}
We note that in the case where $h_1 > h_2 = \dots = h_d$, the matrices $F$ and $\tilde{H}$ coincide up to row and column permutations and 
thus have the same spectrum.

\subsubsection{Decoder message representation}
We note that the message-passing algorithm described above cannot 
be implemented with finite resources.
To implement the decoder in practice, two distinct implementation strategies exist in the classical LDLC literature.

The first, due to Sommer~\emph{et al.}, is often referred to as the \emph{quantized decoder} and represents each message as a discrete vector that corresponds to the probability density on a finite interval with finite resolution. 
We describe this implementation in more detail in \appref{app:quantized_decoder}.
While the complexity of the decoder is still linear in the number of variable nodes, the choice of message representation incurs a large prefactor that depends on the resolution and the range covered.

The second strategy, known as \emph{Gaussian mixture decoders}~\cite{kurkoski_message-passing_2008}, is to represent messages parametrically through the amplitude, mean and variances of the Gaussians in that message, see \appref{app:gaussian_mixture_decoder} for more details.
While this approach sounds at first to have significantly lower overhead, the number of Gaussian components would exponentially increase with the iteration number and thus the mixture must be truncated to a finite set.
Different strategies in choosing this finite subset result in differences in the variable-node update of the algorithm and affect the complexity and accuracy of the decoder.
We have implemented and tested various such strategies and describe them in more detail in \appref{app:gaussian_mixture_decoder}.
In most cases, we employ local 
list sphere decoding~\cite{agrell_closest_2002} 
at each variable node to select Gaussians for the variable node 
update, 
following the proposal of Wang and Mow~\cite{wang_efficient_2023}.
Our implementation of message-passing algorithm is avaible in the open-source Julia package \textsc{LatticeDecoder.jl}~\cite{julia_lattice_decoder}.

\subsection{Performance of classical low-density lattice codes}
The classical decoder succeeds if  $\hat{\xx} = \xx$. 
A failure in the classical decoder can be due to either 
\begin{enumerate}
    \item  $\bar{\vect{a}}\neq\bar{\vect{b}}$, that is, the decoder does not find the closest lattice point to the received signal.%
    \item $H\vect{x}\neq\argmin_{\vect{b}\in\mathbb{Z}^n}\|\vect{y}-G\vect{b} \|$ that is, the received signal is not in the Voronoi cell of the codeword 
    that has been sent. 
    This situation corresponds to a non-correctable error happening.
    We expect this to be more likely as the noise strength increases.
\end{enumerate}
Poltyrev~\cite{poltyrev_coding_1994} %
has shown that it is possible to derive exponential upper and lower bounds 
on the decoding error probability of lattice codes over the AWGN channel.
Based on this analysis, he also defined a generalized notion of channel capacity for lattices, analogous to the capacity of linear codes over the binary symmetric channel~\cite{shannon_mathematical_1948}.

The \emph{capacity} of the AWGN channel is the largest noise variance $\sigma^2$ for which a \emph{maximum-likelihood} (ML) decoder can recover the transmitted lattice point with 
arbitrarily low probability of 
error. 
For an $n$-dimensional lattice $\Lambda$ defined by a generator matrix $G \in \mathbb{R}^{n\times n}$, lattices exist in the asymptotic limit $n\to \infty$ that satisfy this condition if and only if
\begin{align}
    \label{eq:poltyrev_capacity_bound}
    \sigma^2 \leq \frac{|\det G|^{2/n}}{2\pi e},
\end{align}
where $|\det G|$ is the volume of the Voronoi region of the lattice, quantifying lattice density.
As a result, it is common to consider \emph{normalized lattices}, for which $|\det G|^{1/n} = 1$.
Under this assumption, the channel capacity in terms of the standard deviation is approximately
\begin{align}
    \label{eq:classical_ldlc_capacity}
    \sigma_C \approx 0.2420.    
\end{align}

A standard way to represent decoder performance in the classical LDLC community is via the \emph{distance from capacity}, measured in decibels. 
For an AWGN channel with noise variance $\sigma^2$, the distance from capacity is defined as
\begin{align}
\mathrm{SNR}_C\,[\mathrm{dB}] = -10 \log_{10} \left( \frac{\sigma^2}{\sigma_C^2} \right).    
\end{align}
Decoder performance is also commonly quantified using the \emph{symbol error rate} (SER), defined as the proportion of symbols, that is, integer lattice coordinates, in a transmitted codeword that are incorrectly decoded.
It is therefore a measure of how close the decoder's output is to the correct lattice point, not just whether it is correct.

\subsection{Multi-mode Gottesman-Kitaev-Preskill codes}
In this section, we review the Gottesman-Kitaev-Preskill codes first proposed in Ref.~\cite{GKP} using the lattice formalism as introduced in Ref.~\cite{Conrad_2022}. 
We will denote with $\hat{a}=(\hat{q}+i\hat{p})/{\sqrt{2}}$ the bosonic annihilation operator, with $\hat{q},\ \hat{p}$ position- and momentum-like operators, respectively. 
\emph{GKP codes} are defined from a set of stabilizer generators comprising a set of commuting, multi-mode displacement operators~\cite{Conrad_2022}, 
\begin{align}
    D(\vect{\xi}) &= \prod_k \exp\left(
    -i \sqrt{2\pi}\xi_p^{(k)}\hat{p}_k + i\sqrt{2\pi}\xi_q^{(k)}\hat{q}_k 
    \right)
    \nonumber
    \\
    &= \exp\left(
    -i \sqrt{2\pi} \vect{\xi}^TJ\vect{\hat{x}}
    \right),
    \label{equ:GKP_stab}
\end{align} with 
$\vect{\hat{x}} = (q_1, \dots, q_n, p_1, \dots, p_n)^T$
and $J$ the symplectic form
reflecting the canonical commutation relations\footnote{Note that other authors consider an alternative coordinate ordering that is equally common, that is, $\vect{\hat{x}} = (q_1, p_1,  \dots, q_n, p_n)^T$.}
\begin{equation}
    J =\begin{pmatrix}
        0 & 1 \\ -1 & 0
    \end{pmatrix}\otimes  \mathbb{I}_n  .
\end{equation}
The \emph{Weyl-commutation relation}~\cite{weyl1950theory} 
(which actually constitutes the rigorous formulation of the canonical commutation
relations for unbounded operators)
reads
\begin{equation}
    D(\vect{\xi}_k)D(\vect{\xi}_l) = e^{-i2\pi\vect{\xi}_k^TJ\vect{\xi}_l}D(\vect{\xi}_l)D(\vect{\xi}_k).
\end{equation} Therefore, to ensure commutation, one has to require 
\begin{equation}
    \vect{\xi}_k^TJ\vect{\xi}_l \in \mathbb{Z} \quad\forall k, l \in \{1, \dots,  2n\}.
    \label{equ:commutation_relation}
\end{equation} 
Any set of $2n$ displacement operators that satisfy this condition and whose arguments are linearly independent can be taken to define a GKP code with finite logical dimension~\cite{GKP, Conrad_2022}.

A set of displacement operators generates the stabilizer group of a GKP code through composition
\begin{equation}
    \mathcal{S} = \langle D(\vect{\xi}_1), \dots,  D(\vect{\xi}_{2n})\rangle,
\end{equation} with $\{\vect{\xi}_i\}_{i=1}^{2n}$ linearly independent. 
From the relation $D(\vect{\xi}_k)D(\vect{\xi}_l)=e^{-i\pi\vect{\xi}_k^TJ\vect{\xi}_l}D(\vect{\xi}_k+\vect{\xi}_l)$, one deduces that $\mathcal{S}$ is isomorphic to a lattice with generator matrix~\cite{Conrad_2022}
\begin{equation}
    M = \begin{pmatrix}
        \vect{\xi}_1^T \\
        \vdots\\
        \vect{\xi}_{2n}^T \\
    \end{pmatrix}.
\end{equation} 
The lattice is the set of integer linear combinations of basis elements
\begin{equation}
\mathcal{L}(M) = \{\vect{\xi} \in \mathbb{R}^{2n} \mid \vect{\xi}^T = \vect{a}^T M, \vect{a} \in \mathbb{Z}^{2n}\}.
\end{equation}
Clearly, more than one basis corresponds to the same lattice (although choosing a different basis might require allowing some stabilizer generators to act as $-\mathbb{I}$ on the code space~\cite{Conrad_2022}).
We will nevertheless denote by $\mathcal{L}(M)$ the lattice stabilized by displacements with the rows of $M$ as arguments, and by $\mathcal{C}(M)$ the respective code space. 
Eq.~\eqref{equ:commutation_relation} can be written compactly as
\begin{equation}
    A := MJ M^T \in  \mathbb{Z}^{2n \times 2n}, \label{eq:integerGram}
\end{equation} 
that is, as the condition that the symplectic Gram matrix $A$ associated with the generator $M$ has only integer entries.
Logical Pauli operators are defined as the displacement operators that commute with the stabilizer group, i.e., the \emph{normalizer} group of $\mathcal{S}$~\cite{GKP, Conrad_2022, Lin_2023}. 
For a GKP code, this is isomorphic to the symplectic dual 
lattice $\mathcal{L}(M^\perp)$
according to
\begin{equation}
    \mathcal{L}(M^\perp) = \{\vect{\xi}^\perp \in \mathbb{R}^{2n} \mid (\vect{\xi}^\perp)^T J\vect{\xi} \in \mathbb{Z} \;\; \forall \vect{\xi} \in \mathcal{L}(M) \}.
\end{equation} 
and $M^\perp $ can be obtained from~\cite{Lin_2023}
\begin{equation}
    M^\perp = J A^{-1}M.
    \label{equ:Mperp_from_M}
\end{equation} 
The distance $\Delta$ of a GKP code is defined as the Euclidean length of the shortest non-trivial logical operator~\cite{Conrad_2022}, i.e,
\begin{equation}
    \Delta(\mathcal{L}(M)) := \min_{\vect{x}\in\mathcal{L}(M^\perp)\setminus \mathcal{L}(M)} \|\vect{x}\|.
\end{equation} 
Finally, the logical dimension $K$ encoded by the GKP code is given by~\cite{Conrad_2022}
\begin{equation}
    K = |\det(M)|.
\end{equation}

\subsection{Minimum energy decoding of GKP code \label{ssec:gkp_med}}

In this section, we review the \emph{minimum-energy decoding} (MED) strategy for general GKP codes. 
Let us consider an $n$-mode GKP code. 
We assume a common idealized noise model of \emph{Gaussian random noise} (GRN), also known as AWGN in the classical literature,
\begin{equation}
    \vect{\xi} %
    \mapsto 
    \vect{\xi}' 
    \coloneqq
    \vect{\xi} + \vect{e},
\end{equation} 
where $\vect{e}^T = (e_q^{(1)}, e_q^{(2)}, \dots,  e_p^{(1)}, e_p^{(n)}) \sim_{iid} \mathcal{N}(0, {\sigma}^2)$.

The displacement error $\vect{e}$ is unknown; the only accessible information is the syndrome obtained from stabilizer measurements. 
Since the GKP stabilizers, defined in Eq.~\eqref{equ:GKP_stab} as
\begin{equation}
    S_i = \exp\Big(-i \sqrt{2\pi} \, \vect{\xi}_i^T J \vect{\hat{x}}\Big),
\end{equation}
commute, the corresponding observables 
$-\sqrt{2\pi}\, \vect{\xi}_i^T J \vect{\hat{x}}$
can be measured simultaneously modulo $2\pi$. 
The resulting syndrome is therefore
\begin{equation}
    \vect{s} = -\sqrt{2\pi}\, M J \vect{e} \;\; \text{mod } 2\pi,
\end{equation}
where the rows of $M$ correspond to the generating vectors $\vect{\xi}_i$ of the stabilizer group.
Applied elementwise, this modulo operation implies that the displacement error and syndrome are related via $\vect{e} = (-1/\sqrt{2\pi})(MJ)^{-1}(\vect{s} + 2\pi\vect{a})$ for some integer vector $\vect{a}$ ~\cite{Lin_2023}. 
Then by using Eq.~\eqref{equ:Mperp_from_M} we have
\begin{align}
\vect{e} &= -\frac{1}{\sqrt{2\pi}}J^{-1}(J M^T A^{-1})(\vect{s} + 2\pi\vect{a})\\
\nonumber
&= \frac{1}{\sqrt{2\pi}}(J M^{\perp})^T(\vect{s} + 2\pi\vect{a})\\
\nonumber
&= \vect{\eta}(\vect{s}) - \sqrt{2\pi}(M^{\perp})^T\vect{b},
\nonumber
\end{align}
where we define $\vect{b} %
\coloneqq J\vect{a}$ and
\begin{equation}
\vect{\eta}(\vect{s}) %
\coloneqq
\frac{1}{\sqrt{2\pi}}(J M^{\perp})^T\vect{s}.
\end{equation} 
The quantity $\vect{\eta}(\vect{s})$ is sometimes called the pure error; it has the same syndrome as $\vect{e}$.

The syndrome reveals the displacement only modulo the logical operator lattice $\sqrt{2\pi}M^{\perp}$. The optimal decoder, known as the \emph{maximum-likelihood decoder} (MLD), maps $\boldsymbol{s}\mapsto \boldsymbol{e}^*$ such that $\boldsymbol{e}^*$ maximizes the probability that $D(\vect{e})D(-\vect{e}^*)\in\mathcal{S}$.
A common approximation, accurate in the low-error regime, is to choose the single most probable displacement $\vect{e}^*$ consistent with the measured syndrome $\vect{s}$, rather than the most likely coset. 
This procedure is called minimum energy decoding, and it reduces to solving~\cite{Conrad_2022} the equation
\begin{equation}
\vect{b}^* = \arg\min_{\vect{b} \in \mathbb{Z}^{2N}} |\vect{\eta}(\vect{s}) - \sqrt{2\pi}(M^{\perp})^T\vect{b}|.
\label{equ:getting_shortest_displace}
\end{equation}
This finds the shortest displacement whose syndrome equals the measured syndrome $\vect{s}$, which is consistent with the assumption that errors follow a normal distribution centered at zero.
The recovery operation is $D(-\vect{e}^*)$ with $\vect{e}^* = \vect{\eta}(\vect{s}) - \sqrt{2\pi}(M^{\perp})^T\vect{b}^*$.
Solving Eq.~\eqref{equ:getting_shortest_displace} corresponds to solving the \emph{closest vector problem} for the dual lattice. 
This provides the link with classical lattice decoding algorithms.

\section{Defining GKP low-density lattices}

Our goal in this section is to modify the construction of classical LDLCs so that the parity check matrix satisfies Eq.~\eqref{eq:integerGram}. 

\subsection{Magic square quantum low-density lattice codes}
The first construction we describe is somewhat trivial and follows the construction of magic-square LDLCs in the original paper by Sommer~\emph{et al.}~\cite{ldlc}. 
There, the authors restrict 
their attention to $d$-regular LDLCs where the parity check matrix $H$ has the same $d$ non-zero entries in each row and column up to a random sign, namely $h_0 = \pm 1,\ h_1,\ldots,  h_{d-1} = \pm \frac{1}{\sqrt{d}}$. 
It is then trivial to satisfy Eq.~\eqref{eq:integerGram} by choosing $d=z^2$, $z\in\mathbb{N}$ and defining $M = \sqrt{d}H$. 
The multiplication by $\sqrt{d}H$ has two side-effects: first, the logical dimension becomes exponentially large in the number of modes since $\det(M) = d^{N}\det(H)$. 
Secondly, the distance typically becomes very small. 
Intuitively, this is a consequence of packing more logical dimensions in the same number of modes: under CVP decoding, displacement errors can generate logical errors if they land outside the Voronoi cell of the (symplectic) dual lattice. But this is generated by $M^\perp = \left(JM^T\right)^{-1}$ and so its volume scales as $\sim d^{-N}$. 
Put another way, re-scaling the generator has the same effect as re-scaling the noise; therefore these trivial codes generally have high logical error rates at moderate physical noise strength.
 
\subsection{Quantum almost-low-density lattice codes}\label{sec:dim_red}
To circumvent the drawbacks of the previous construction, we set out to reduce the logical dimension. 
We describe here a strategy to start from a magic square GKP-LDLC and obtain a GKP qubit with higher distance by appending to a sparse generator $M$ only two dense rows, provided its generator has a specific canonical form. Recall that every GKP code admits a generator $M_\mathrm{can}$ such that~\cite{GKP,Conrad_2022}
\begin{equation}
M_\mathrm{can}JM_\mathrm{can}^T = \begin{pmatrix}
    0 & -1 \\ 1 & 0
\end{pmatrix}\otimes D
\end{equation}
with $D$ a diagonal matrix with positive integer diagonal entries $D_{j,j}=d_j$.
This canonical form is unique if one requires the divisibility chain $d_1\mid d_2\mid \cdots\mid d_n$~\cite{Burchards2025fiberbundlefault}; the $d_j$ are then the \emph{invariant factors} of the symplectic Gram matrix $A=MJM^T$, and their product $\prod_j d_j = \sqrt{\det A}=K$ is the logical dimension. Each factor $d_j$ is the logical dimension carried by the $j$-th mode, so the largest factor $\max_j d_j$ sets the maximum mode-wise (``local'') logical dimension. Perhaps of separate interest is the fact that this maximum can be lowered within a fixed equivalence class. This is discussed in~\appref{app:invariants}.
Our strategy for dimensionality reduction is detailed here in the case that $M_\mathrm{can}$ has a specific canonical form, and runs as follows.
\begin{enumerate}
        \item Start from a code as in the previous section with $M = \sqrt{d}H$.
        \item Check that $M$ has 
        standard form $M_\mathrm{can}= U_\mathrm{can}MU_\mathrm{can}^T =i \sigma_Y \otimes D$, where $\sigma_Y$ is the Pauli $Y$ matrix, $D = \mathbb{I}\oplus ( 2 \ell)$ for some integer $\ell$ and unimodular matrix $U_\mathrm{can}$. Otherwise, pick another code.
        \item Call $\vect{u} = M_\mathrm{can}(n);\ \vect{v}= M_\mathrm{can}(2n)$, the $n$th and $2n$th rows of $M_\mathrm{can}$, and define $\tilde{\vect{u}} = \vect{u}/k_1$, $\tilde{\vect{v}} = \vect{v}/k_2$ for integers 
        $k_1\times k_2 = \ell$.
        \item Define the generator of the new code \begin{equation}
            \tilde{M} = \begin{pmatrix}
                M_\mathrm{can}(1)\\
                \vdots\\
                M_\mathrm{can}(n-1)\\
                \tilde{\vect{u}}\\
                M_\mathrm{can}(n+1)\\
                \vdots\\
                M_\mathrm{can}(2n-1)\\
                \tilde{\vect{v}}
            \end{pmatrix}.
        \end{equation}
\end{enumerate} By construction, the code $\mathcal{C}(\tilde{M})$ (whose stabilizer is isomorphic to the lattice generated by rows of $\tilde{M}$) encodes a single qubit, since $\det\tilde{M}=2$. 

Note that here we assumed that $M_\mathrm{can}$ is in canonical form, which is not sparse in general. However, the integer span of the rows of $\tilde{M}$ is the same as the integer span of the rows of $M$ complemented with $\tilde{u},\ \tilde{v}$.
Therefore, the single qubit we found is stabilized by displacements corresponding to the rows of 
$   M_\mathrm{ext} =\begin{pmatrix}
        M^T|\tilde{\bs{u}}^T|\tilde{\bs{v}}^T
    \end{pmatrix}^T$, 
which are all sparse except for the 
last two.
    
Logical operators for the new qubit are readily 
found as $X_L=D\left(\frac{\tilde{\bf{u}}}{2}\right)$ and $Z_L=D\left(\frac{\tilde{\bf{v}}}{2}\right)$. 
Their commutation is verified from the symplectic product of their arguments, which is $1/2$ by construction. 
To find the distance of the code, we need to find short representatives of the logical operators.
This can be done for small instances by solving
CVP.
The complete reduction, from the trivial GKP-LDLC generator to the single-qubit code and its distances, is summarized as pseudocode in \algref{alg:dim_red}. In practice every step (canonical form, factor splitting, and coset reconstruction) can be carried out in exact rational arithmetic, and the CVP is preceded by an exact lattice reduction, as detailed in \appref{app:dim_red}.

\begin{algorithm}[!htb]
\SetAlgoLined
\KwIn{GKP-LDLC generator $M \in \mathbb{Z}^{2n\times 2n}$, e.g.\ $M=\sqrt{d}H$}
\KwOut{Single-qubit generator $\tilde{M}$, logical operators $X_L,Y_L,Z_L$ and distances $\Delta_X\le\Delta_Z\le\Delta_Y$; or \textsc{reject}}

$A \gets M J M^{T}$\;
Compute the integer symplectic canonical form $F = C A C^{T} = J_2\otimes D$ with $C$ unimodular, and read off the invariant factors $D=\mathrm{diag}(d_1,\dots,d_n)$, $d_1\mid\cdots\mid d_n$\;

\tcp{Reducibility to a single qubit}
\eIf{$d_1=\cdots=d_{n-1}=1$ \textbf{and} $d_n$ is even}{
    $\ell \gets d_n/2$\;
}{
    \Return \textsc{reject}\;
}

$M_\mathrm{can} \gets C M$\;
Choose $k_1,k_2$ with $k_1 k_2 = \ell$ and $k_1$ the divisor of $\ell$ closest to $\sqrt{\ell}$\;
$\tilde{\vect u} \gets M_\mathrm{can}(n)\,/\,k_1$\;
$\tilde{\vect v} \gets M_\mathrm{can}(2n)\,/\,k_2$\;
Form $\tilde{M}$ from $M_\mathrm{can}$ by replacing row $n$ with $\tilde{\vect u}$ and row $2n$ with $\tilde{\vect v}$ \tcp*{$|\det\tilde{M}|=2$}
$X_L \gets \tilde{\vect u}/2$;\quad $Z_L \gets \tilde{\vect v}/2$;\quad $Y_L \gets X_L + Z_L$\;

\For{$P \in \{X_L,Y_L,Z_L\}$}{
    $\vect{s}_P \gets$ shortest representative of $P$ modulo the stabilizer lattice $\mathcal{L}(\tilde{M})$ \tcp*{solve CVP}
    $\Delta_P \gets \lVert \vect{s}_P \rVert$\;
}
Relabel $(X,Z,Y)$ so that $\Delta_X\le\Delta_Z\le\Delta_Y$\;

\Return $\tilde{M},\; X_L,Y_L,Z_L,\; \Delta_X,\Delta_Z,\Delta_Y$\;
\caption{Single-qubit dimensionality reduction of a GKP-LDLC}
\label{alg:dim_red}
\end{algorithm}

The procedure above can be generalized to accommodate different logical dimensions and extended to the case where the matrix $D$ in the canonical form has more than one non-unit entry. In that regime the invariant factors can be redistributed across modes without leaving the Gaussian-unitary equivalence class of the code, which can be used to lower the largest local dimension $\max_j d_j$ and thereby improve the normal-form distance; we detail this in \appref{app:invariants}. Algorithm~\ref{alg:dim_red} is implemented in the package \textsc{SymplecticGKP.jl}~\cite{julia_symplectic_gkp}.

\subsection{Distance and logical error probability of the reduced codes}\label{sec:voronoi}

\begin{figure*}
	\centering
    \includegraphics{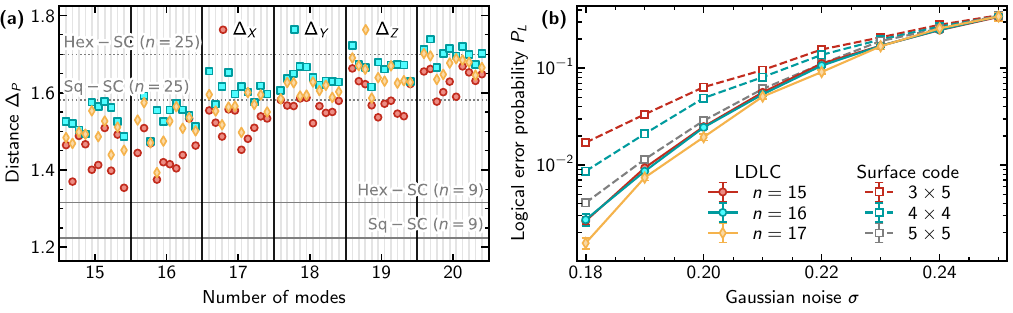}
	\caption{Dimension-reduced GKP-LDLC codes compared to surface-code baselines.
		\textbf{(a)} Euclidean distances of randomly generated GKP-LDLC codes after reducing the logical dimension to a single qubit as explained in \secref{sec:dim_red}. 
        For each instance we plot the three logical distances $\Delta_X$ (circles), $\Delta_Y$ (squares) and $\Delta_Z$ (diamonds); all three are inequivalent because the reduced code is no longer CSS. 
        Note that which displacements correspond to $X,Y,Z$ logicals is arbitrary, so here we sorted them for increasing distance for visual clarity. 
        Instances are grouped by the number of modes $n$. 
        Horizontal lines mark the distances of rotated surface codes concatenated with the single-mode square GKP code 
        and hexagonal GKP code. 
		\textbf{(b)} Logical error probability $P_L$ of the reduced GKP-LDLC codes under an isotropic Gaussian displacement channel of standard deviation $\sigma$, estimated by Monte-Carlo nearest-point (CVP) decoding on the dual lattice. 
        Solid lines with filled markers are GKP-LDLC codes; dashed lines with open squares are rotated surface codes concatenated with the hexagonal single-mode GKP code, matched by mode number.
        }
	\label{fig:reduced_codes}
\end{figure*}

We performed numerical experiments and found that already at relatively small sizes ($n\sim 15$) the reduced codes have a higher distance than the distance-$3$ (rotated) surface code concatenated with the single-mode square or hexagonal GKP code, and that for $n\gtrsim 20$ they are comparable to the distance-$5$ ($n=25$) surface-code baselines.
These reference distances are computed from the relation $n = d^2$ (more generally $n = d_xd_z$ for a rotated $d_x\times d_z$ patch), valid for rotated surface codes, and the relation $\Delta_\mathrm{conc}=\sqrt{d}\,\Delta_\mathrm{loc}$~\cite{Conrad_2022,conradPhD}, valid for CSS codes concatenated with the square or hexagonal single-mode GKP codes.
Here, $\Delta_\mathrm{conc}$ is the Euclidean distance of the concatenated GKP code, while $\Delta_\mathrm{loc}$ is the Euclidean distance of the inner level GKP code.

The distances of ten random instances per mode number are shown and compared to the surface code in \sfigref{fig:reduced_codes}{a}, with the accessible sizes limited by the computational cost of performing CVP for larger instances.

The minimum distance is a worst-case quantity: it controls only the leading term of the logical error probability in the low-noise limit. A more informative figure of merit is the \emph{logical error probability} $P_L$ under an isotropic Gaussian displacement channel of variance $\sigma^2$, decoded by a nearest-point (minimum-energy) decoder. 
Geometrically, $P_L$ is the Gaussian measure of the complement, within the dual lattice $\mathcal{L}(M^\perp)$ [Eq.~\eqref{equ:Mperp_from_M}], of the Voronoi cell of the origin: a displacement is corrected when it decodes back onto the stabilizer lattice $\mathcal{L}(M)$, and counted as a logical error otherwise. 
Unlike the distance, $P_L$ integrates over all directions in phase space, weighted by their probability, and thus accounts for the many near-degenerate facets of the high-dimensional Voronoi cell rather than a single shortest vector.

We estimate $P_L$ by Monte Carlo, drawing Gaussian displacements and decoding each by solving CVP on the dual lattice; the logical class of the outcome is read off by its symplectic pairing with $X_L$ and $Z_L$. \sfigref{fig:reduced_codes}{b} compares the reduced GKP-LDLC codes at $n=15,16,17$ with rotated surface codes concatenated with the hexagonal single-mode GKP code, matched mode-for-mode ($3\times5$, $4\times4$ and $5\times5$ patches, the last using $n=25$ modes). 
Only the hexagonal case is shown as it consistently outperforms the square code in this measure. 
The qLDLC codes improve upon their matched surface-code partners, and the advantage widens as $\sigma\to0$, consistent with their larger distance; strikingly, at the lowest noise studied a $15$--$17$-mode qLDLC code beats even the $25$-mode surface code.

We stress that this comparison concerns the intrinsic quality of the codes, not a practical decoder. Evaluating $P_L$ requires solving a CVP for every displacement, which is efficient only for the well-conditioned surface-code dual lattices. 
The GKP-LDLC dual lattices are, by contrast, highly skewed (with orthogonality defect $\sim 10^{11}$),\footnote{
The orthogonality defect of a lattice basis $\{\vect b_i\}$ is $\delta=\prod_i\lVert\vect b_i\rVert/\operatorname{covol}(\mathcal L)\ge 1$, where the covolume $\operatorname{covol}(\mathcal L)=\sqrt{\det(BB^T)}$ (with $B$ the generator whose rows are the $\vect b_i$) is a basis-independent invariant. 
Equality $\delta=1$ holds if and only if the $\vect b_i$ are mutually orthogonal, so a large $\delta$ signals a strongly non-orthogonal (skewed) basis whose fundamental cell is a thin, sheared parallelepiped. 
For such a basis, closest-vector decoding by simple coordinate rounding fails, and exact CVP requires a costly enumeration whose size grows with $\delta$ and the dimension.} so exact CVP decoding is feasible only up to $n\approx 16$--$17$. 
Reaching the mode numbers of practical interest ($n\sim 50$) therefore would require an efficient approximate decoder. 
The obvious candidates are the analog message-passing decoders discussed in \secref{sec:mp_decoding}. Unfortunately, the introduction of even two dense checks introduces too many short cycles in the decoding graph for these algorithms to perform well, and a simple way was not found to adapt them to this case. 
Instead, in Sec.~\ref{sec:mp_decoding} we turn to GKP codes concatenated with qubit-level LDPC codes, which can be constructed to avoid short cycles.

\subsection{qLDLCs from qudit CSS codes}

We conclude by mentioning a strategy to construct GKP codes with sparse parity check matrices starting from qu\emph{dit} low-density-parity-check codes.
It is possible to define qLDPC codes over qudits of prime-power dimension~\cite{Spencer2026quditlowdensity}. 
In particular, CSS constructions exist where the CSS condition reads $H_XH_Z^T=0\mod p^l$, with $p$ the prime power and $l\in \mathbb{N}$, $p^l$ being the 
qudit's dimension. 
If we define $H = H_X\oplus H_Z$ it is relatively straightforward to complete $H$ to the generator of a GKP code as 
\begin{equation}
    M = \begin{pmatrix}
        p^{-l/2}H\\ p^{l/2}\mathbb{I}
    \end{pmatrix}
\end{equation} 
which is a rescaled version of a construction A lattice~\cite{ConwaySloane:1999,Conrad_2022}. 
It can be verified immediately that 
\begin{equation}
    MJM^T = \begin{pmatrix}
        0 & p^{-l}H_XH_Z^T & 0 & H_X \\
        -p^{-l}H_ZH_X^T & 0 & -H_Z & 0 \\
        0 & H_Z^T & 0 & p^l\mathbb{I} \\
        -H_X^T & 0 & -p^l\mathbb{I} & 0
    \end{pmatrix}
\end{equation} 
which has integer entries by construction. 
Furthermore, $M$ is of full rank, and it inherits the sparsity of the parity checks of the underlying qudit code.

\section{Decoding concatenated codes with analog message-passing decoders}\label{sec:mp_decoding}
Another natural direction of investigation is to apply the analog MP decoder to previously known concatenated GKP-Stabilizer codes. 
For testing purposes, we focus on the square code concatenated with the \textit{repetition code}. 
As an example of a fully quantum code, we then turn to small instances of simplex codes~\cite{panteleev_degenerate_2021, webster_explicit_2025}. 

In the quantum setting, the structure of low-density lattice codes is typically irregular, so a rigorous extension of Sommer's convergence analysis for mean values is not straightforward.
It turns out that blindly applying analog message-passing decoders to concatenated lattice codes does not yield good results.
At this point, we do not yet understand how to modify the decoder to achieve highly accurate decoding across a large class of codes.
Nevertheless, there exist concatenated codes for which the decoder displays a threhold. 
We also note that similar to its discrete-variable counterpart, details of the decoding algorithm such as the node update schedule and the variable node iteration seem to impact performance quite dramatically. 
We focus on giving a few examples that might lead to further explorations. 
We begin by outlining an additional step we can run after the message-passing decoder, which is reminiscent of ordered statistics decoding for qubit codes.

\subsection{Reprocessing and local search}
In  the \emph{discrete-variable} (DV) setting, 
\emph{belief propagation} (BP) and related message-passing decoders are known to converge exactly only on loopless (tree-like) factor graphs. 
In practice, however, most relevant LDPC codes contain many short loops. 
As a result, BP often fails to converge or becomes trapped in local minima, leading to error floors. 
To mitigate this, Fossorier and Lin~\cite{fossorier_soft-decision_1995} 
have introduced \emph{ordered statistics decoding} (OSD), a post-processing step that refines the possibly invalid or suboptimal solutions returned by BP, thereby significantly reducing error floors in classical LDPC decoding. 
Inspired by this idea, we develop an analogous OSD-style postprocessor for LDLCs, adapting the core principle of targeted local search in the most uncertain components of the solution. 

We begin by giving a high-level explanation of the idea. 
For this, recall that the hard decision vector is given by $\hat{\aa} = \lfloor H\tilde{\bar{\xx}} \rceil \in \mathbb{Z}^n$.
The post-processing routine searches for an improved solution for a restricted set of coordinates determined by the index set $S$, that is,
\begin{align}
    \hat{\tilde{\aa}} = \hat{\aa} + V_S \vect{u},
\end{align}
where $\vect{u} \in \mathbb{Z}^w$ with $w = \lvert S \rvert$ is the size of the index set and matrix $V_S \in \mathbb{Z}^{n \times w}$ performs the natural injection (embedding) of $\mathbb{Z}^w \hookrightarrow \mathbb{Z}^n$ by placing entries in the coordinates indexed by $S$. 
The corresponding lattice point then is
\begin{align}
    \hat{\vect{x}} = G \hat{\tilde{\aa}} = G  \hat{\aa} + G V_S \vect{u}.
\end{align}
Upon defining $B = G V_S$, the post-processing problem is
\begin{align}
    \label{eq:cvp_local_search}
    \min_{u \in \mathbb{Z}^w} \lvert \lvert (\vect{y} - \tilde{\vect{x}}) - B \vect{u} \rvert \rvert^2,
\end{align}
which is a CVP problem over the lattice basis $B$, but only in $0 < w \leq n$ dimensions and thus can be solved significantly faster.

Instead of solving the CVP problem exactly, we consider solving it approximately.
To determine the index set $S$, first we select the $w$ positions with the lowest reliability $r_k$ given by 
\begin{align}
    r_k = \left[ \min_{m \in \mathbb{Z}}\lvert (H \tilde{\xx})_k  - m\rvert \right]^{-1},
\end{align}
although one may choose an alternative reliability metric such as the $w$ largest posterior variances.
Then, given the matrix $B = G V_S$, we perform a lattice basis reduction of $B$ to obtain a short, nearly orthogonal lattice basis using the Lenstra–Lenstra–Lovász algorithm~\cite{lenstra_factoring_1982}.
This allows computing a fast heuristic solution to the CVP~\eqref{eq:cvp_local_search} as the integer least-square approximation
\begin{align}
    \vect{u}_{\mathrm{init}} =  \lfloor (B^T B)^{-1} B^T (\vect{y} - \tilde{\vect{x}}) \rceil. 
\end{align}
In practice, this corresponds to the first step of the Babai nearest-plane algorithm~\cite{babai_lovasz_1986}, and provides a good initialization for the subsequent search.
For simplicity, we implement a bounded search restricting each coordinate $u_i$ to $2L + 1$ integer values, that is, $u_i \in \{ u_{\mathrm{init}, i} - L , \dots, u_{\mathrm{init}, i} + L \}$ which is efficient enough for small $w$.
For larger $w$ it is likely more efficient to run a sphere decoder~\cite{agrell_closest_2002} centered at $\vect{u}_{\mathrm{init}}$ with radius $R = \lvert \lvert (\vect{y} - \tilde{\vect{x}}) - B \vect{u}_{\mathrm{init}}\rvert \rvert$.
In the case of the bounded search, from all the candidates $\vect{u}$, we select the one that minimizes the distance $\lvert \lvert (\vect{y} - \tilde{\vect{x}}) - B \vect{u} \rvert \rvert$ and update $\hat{\aa}$ accordingly.

\subsection{Repetition codes}

\begin{figure}
    \includegraphics{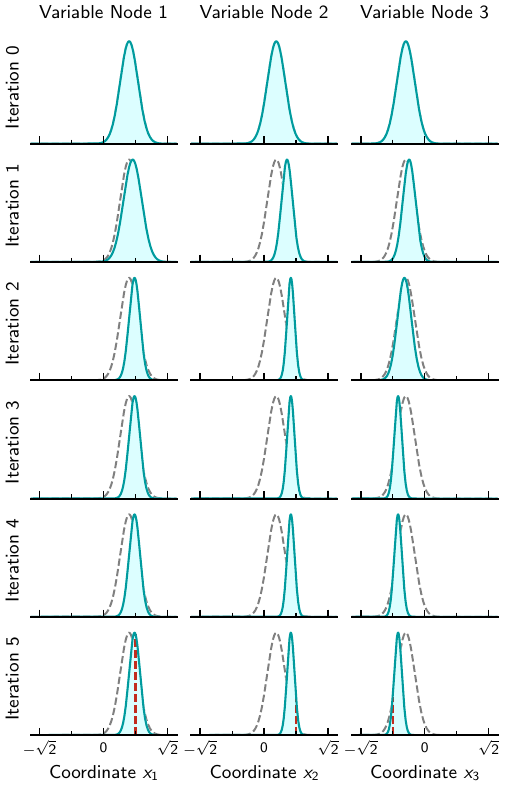}
    \caption{Evolution of messages for the repetition lattice code over five iterations. 
    Blue Gaussians show the mean and variance of the final marginal if 
    it were calculated after the $k\textsuperscript{th}$ iteration.
    The grey dashed line shows the Gaussian with initial mean obtained from the syndrome measurement, and its variance is the channel 
    prior.
    The red vertical dashed 
    lines 
    show the decoded lattice point obtained after the hard 
    decision, see Eq.~\eqref{eq:hard_decision}.
    For visualization 
    purposes, Gaussians are not normalized.
    }
    \label{fig:placeholder}
\end{figure}

As a first example, consider the three-qubit repetition code concatenated with the square GKP code. 
A possible generator matrix is
\begin{equation}
\label{eq:rep_code}
    M_3 = \frac{1}{\sqrt{2}} \left[2\mathbb{I}\oplus \begin{pmatrix}
        2 & 0 & 0 \\ 1 & 1 & 0 \\ 0 & 1 & 1
    \end{pmatrix} \right].
\end{equation} 
This is a CSS code, where the non-trivial qubit-level stabilizers contain only position operators. 
The lattice defining the GKP code splits accordingly into momentum and position sectors. 
Focusing on the position sector, one has a linear decoding graph with three variable nodes and three check nodes. 
A typical evolution of the messages from the Gaussian mixture implementation (see \appref{app:gaussian_mixture_decoder}) of the decoder is shown in Fig.~\ref{fig:placeholder}.

\begin{figure}[ht]
    \includegraphics{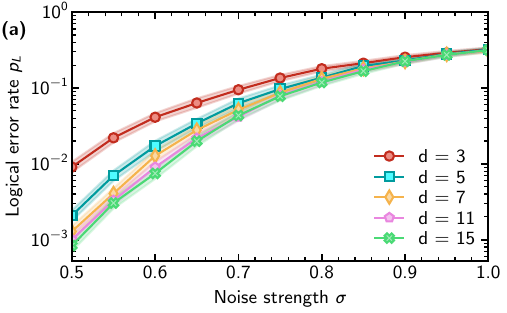}
    \includegraphics{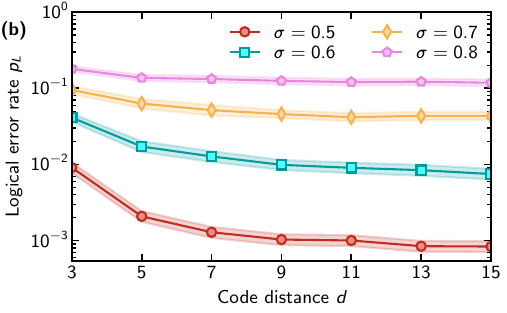}
    \caption{Position sector error rates for the GKP-repetition code of various length using the Gaussian mixture decoder using a serial schedule and the list sphere decoding variable-node update rule with uniform memory $\gamma = 0.15$ over 50 iterations. (a) The estimated logical error rate as a function of the noise strength for distances of the outer repetition code $d$. (b) Error rate as a function of distance for selected physical error strengths.}
    \label{fig:rep_code_thr1}
\end{figure}

A threshold plot and subthreshold behavior are shown in Fig.~\ref{fig:rep_code_thr1} using a serial schedule and the list-sphere decoding update rule for the variable nodes with $50$ iterations and a 
uniform variable-node memory $\gamma = 0.15$. 
While increasing the distance decreases the error rate, the error rate reaches an error floor. 
Running the local search post-processing routine on $d/2$ positions partially alleviates this issue (not shown), but at increasingly higher decoding complexity.

One might find the above result surprising following the usual intuition that message-passing on a tree is exact and thus should work for this example without issues.
However, this argument is too simple as we are not implementing the exact message-passing algorithm described in \secref{sssec:ldlc_decoding}, but instead have implemented variable node updates approximately within the Gaussian mixture framework, see~\appref{app:gaussian_mixture_decoder}.
Indeed, in the classical literature on low-density lattice codes, the case of low degree graphs is typically omitted and convergence is only proven in the, more relevant, case of higher degree magic-square LDLC, see, e.g., Ref.~\cite{liu_efficient_2019}.

\begin{figure}[ht]
        \includegraphics[width=0.48\textwidth]{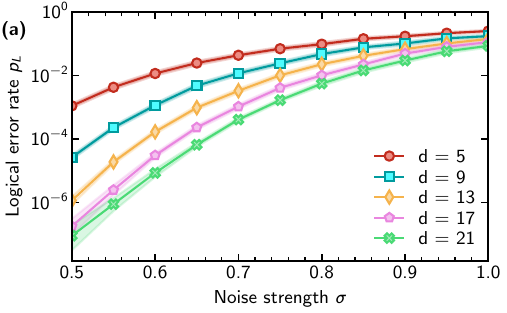}
        \includegraphics[width=0.48\textwidth]{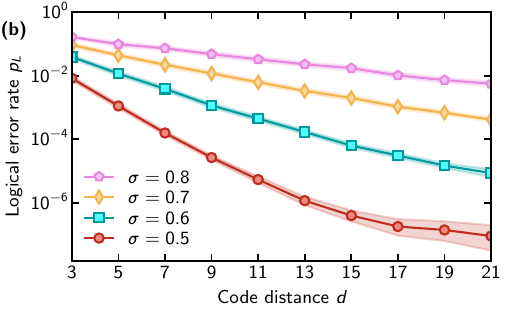}
    \caption{Position sector error rates for the GKP-repetition code using the quantized LDLC decoder using $L = 256$ quantization bins and bin resolution $\Delta = 1/32$ over 50 iterations. (a) The estimated logical error rate as a function of the noise strength for distances of the outer repetition code $d$. (b) Error rate as a function of distance for selected physical error strengths.}
    \label{fig:rep_code_quantized}
\end{figure}

\begin{figure}[!b]
    \centering
    \includegraphics{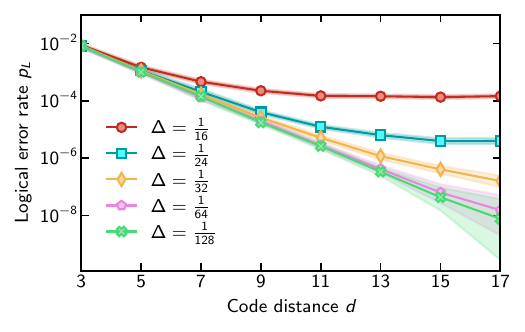}
    \caption{Position sector error rate scaling for the GKP-repetition code using 
    the quantized decoder at $\sigma = 0.5$ with $L = 512$ quantization bins and varying resolution $\Delta$.
    For coarse resolution (large $\Delta$), the decoder develops 
    an error floor, which can 
    be lowered arbitrarily by using 
    a finer resolution (smaller $\Delta$).}
    \label{fig:rep_code_quantization_plot}
\end{figure}

\pagebreak
In \figref{fig:rep_code_quantized} we show the logical error rate performance of the repetition code using our implementation of the quantized decoder of Sommer \emph{et al.}, see~\appref{app:quantized_decoder}, with a resolution $\Delta = 1/32$ over $L = 256$ quantization bins.
Compared to the Gaussian mixture decoder implementation, the quantized decoder implementation shows a higher threshold and has an error floor that is independent of the noise rate $\sigma$ which only becomes apparent at low error rates and high distances, see also~\sfigref{fig:rep_code_quantized}{b}.
In \figref{fig:rep_code_quantization_plot} we show that the error floor in the quantized decoder is again due to the approximate nature of the decoder and it can be removed by increasing the resolution (lower $\Delta$) in the decoder.
Alternatively, post-processing methods can be employed to reduce the error floor.

\subsection{Generalized bicycle codes}
\begin{figure}[!t]
    \centering
    \includegraphics{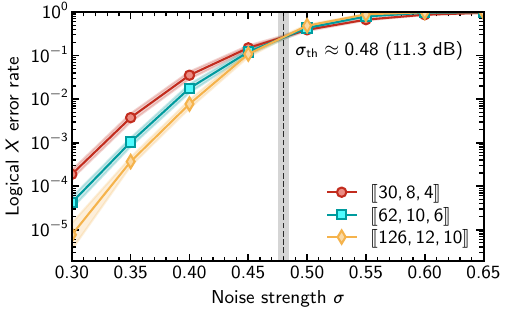}
    \caption{Logical $X$ error rate of weight-6 Hamming-based generalized bicycle codes.
    The figure shows the code capacity noise performance of $\llbracket 30, 8, 4 \rrbracket$, $\llbracket 62, 10, 6 \rrbracket$, and $\llbracket 126, 12, 10 \rrbracket$ generalized bicycle code constructed in Ref.~\cite{webster_explicit_2025}.
    The vertical dashed line indicates the approximate threshold value of $\sigma_{\mathrm{th}} \approx 0.48$.}
    \label{fig:gb_codes_decoder_performance}
\end{figure}

As an example to showcase the performance of the decoder on quantum codes with larger graph degree, we consider small instances of Hamming-based generalized bicycle codes~\cite{panteleev_degenerate_2021}, also known as simplex codes~\cite{webster_explicit_2025}.
\figref{fig:gb_codes_decoder_performance} shows the logical $X$ error rate of the $\llbracket 30, 8, 4 \rrbracket$, $\llbracket 62, 10, 6 \rrbracket$, and $\llbracket 126, 12, 10 \rrbracket$ codes subject to the code capacity noise model characterized by the noise strength $\sigma$.
The performance is achieved by running 50 iterations of the decoder, using a serial update schedule, and using a list-sphere decoding updating schedule for the variable nodes.
We also employ memory $\gamma = 0.15$ in the update rules for increased performance, see \appref{sec:LDLCBP}.
For these instances, no post-processing is applied, demonstrating the good convergence properties of the decoder for the right set of codes.
Notably, the threshold of the code and decoder combination is not significantly behind that of concatenated surface codes~\cite{Lin_2023}.

\section{Quantum logic}

In recent years and months, emphasis in the field of quantum error correction has shifted from studying properties of quantum memories to taking steps to understand how elements of quantum logic can be feasibly and fault-tolerantly implemented.
It should be clear that the codes introduced here readily allow for the implementation of all Clifford operations on the logical level. 
This is briefly pointed out here.
The collection of GKP Clifford gates constitutes a distinguished and structurally significant class of operations acting on the logical subspace of a GKP code. This class is represented by the symplectic automorphism group
\begin{align}
\text{Aut}_{S_\infty}
({\cal L}(M^\perp)) = \text{Aut}_{S}
({\cal L}(M^\perp))\ltimes {\cal L}(M^\perp)
\end{align}
where $\text{Aut}_{S} ({\cal L}(M^\perp))$
denotes the group of symplectic automorphisms of the 
lattice ${\cal L}(M^\perp)$. 
Every logical Clifford operation admits a unique decomposition into (i) a displacement by an element of 
${\cal L}(M^\perp)$ -- referred to as a trivial Clifford operation, as it acts by conjugation on Pauli operators (i.e., displacements in ${\cal L}(M^\perp)$) up to a phase -- and (ii) a symplectic automorphism of 
${\cal L}(M^\perp)$, which maps logical Pauli operators to logical Pauli operators while fixing the identity element. 
This means that every logical Clifford operation can be implemented by a symplectic operation on the physical level, even though the converse is not true. 
That is to say, by construction, fault-tolerant logical Clifford operations are already inbuilt in the lattice-based GKP code construction discussed here.

\section{Discussion and outlook}
In this work, we explore new strategies to construct and decode GKP codes based on techniques imported from the classical coding literature. 
In particular, we build on classical \emph{low-density lattice codes} (LDLC), which were introduced as a lattice version of low-density parity check codes so that they would admit an efficient message-passing decoder. 
As for qLDPC codes, a quantum version of LDLCs would ensure that physical constraints on the number of checks per qubit and qubits per check is constant as the size of the code is increased, making them appealing from a physical standpoint. 
While a naive translation of LDLCs to the quantum domain does not yield good code parameters, we have shown that by reducing the logical dimension, we can find codes whose distance exceeds that of concatenated GKP--surface codes for small numbers of modes (\sfigref{fig:reduced_codes}{a}). 
The advantage is not confined to the minimum distance: evaluated through the full logical error probability under a Gaussian displacement channel, these reduced codes lie below their mode-matched hexagonal GKP--surface-code counterparts, and a $15$--$17$-mode instance already outperforms the $25$-mode ($5\times5$) surface code at low noise (\sfigref{fig:reduced_codes}{b}). 
This dimensionality reduction is achieved by introducing two additional non-sparse checks. 
As a consequence, it is not clear at this stage if the resulting codes are still suitable for decoding through message-passing, and, if so, how the decoder would need to be modified. 
Failure of the current instances is possibly simply due to the instances being too small.
We leave this for future work.

We then studied the performance of a few variants of the message-passing decoder when applied to concatenated GKP-stabilizer codes where 
the outer code has a sparse parity check matrix. 
The upshot of this approach is that a fully analog decoder, if effective, would be preferable to standard approaches based on two-stage decoders typically employed for concatenated GKP codes, where a mode-wise correction is followed by a qubit-level correction, sometimes informed by the analog information gathered at the first stage. 
Our explorations, however, suggest that other conditions, beyond the sparsity of the parity checks, need to be met for the decoder to be effective. 
In the quantum setting, the structure of low-density lattice codes is typically irregular, so a rigorous extension of Sommer's convergence analysis for mean values is not straightforward.
We conducted numerical experiments applying analog message-passing decoders to concatenated lattice codes. 
The results are encouraging for codes whose Tanner graphs have sufficiently high degree: for the weight-$6$ Hamming-based generalized bicycle (simplex) codes the decoder converges without any post-processing and exhibits a pseudo-threshold not significantly behind that of concatenated surface codes (\figref{fig:gb_codes_decoder_performance}). 
Low-degree GKP-repetition codes are more demanding: the approximate Gaussian-mixture decoder develops an error floor (\figref{fig:rep_code_thr1}). 
However, the quantized decoder attains a higher threshold, and its residual floor is a finite-resolution artifact that can be systematically suppressed by refining the quantization (see \figref{fig:rep_code_quantized} and \figref{fig:rep_code_quantization_plot}) or with light post-processing. 
These findings indicate that the effectiveness of the analog decoder is governed by the structure and degree of the underlying Tanner graph rather than by parity-check sparsity alone.
Together, these results highlight the need to develop further understanding of the application of analog message-passing to the decoding problem of quantum codes.

At this point, we have not yet understood how to modify the decoder to consistently achieve highly accurate decoding for dimensionality-reduced GKP-LDLCs or for generic concatenated GKP-qLDPC codes.
Particularly, the convergence requirements of Sommer \emph{et al.}~\cite{ldlc} cannot be applied to the concatenated constructions typically present in the literature, such as construction A lattices.
Therefore, there seem to be two natural avenues for further investigation. 
The first, on the code construction side, would be finding alternative constructions that directly yield CSS-type quantum LDLCs that can fulfill the convergence requirements derived for the classical construction. 
The second, on the decoder side, would be to modify the message-passing decoder to ensure convergence when a few dense checks are introduced or for more generic qLDPCs at the outer concatenation layer.

\acknowledgments
F.~A.\ and T.~H.\ are grateful to Peter-Jan~H.~S.~Derks for introducing them four years ago and thereby initiating this collaboration, and to Joschka Roffe, Erik Agrell and Brian M.\ Kurkoski for useful discussions. T.~H.\ acknowledges the financial support from the Chalmers Excellence Initiative Nano, the Knut and Alice Wallenberg Foundation through the Wallenberg Centre for Quantum Technology (WACQT), the Defence Science and Technologies Group (DSTG) and the Advanced Strategic Capabilities Accelerator (ASCA) through its Emerging and Disruptive Technologies (EDT) Program.
This work has been supported by the Knut and Alice Wallenberg Foundation through the Wallenberg Centre for Quantum Technology via the guest researcher programme. J.~E.~acknowledges funding by the BMFTR (PasQuops, 
QSolid, MuniQC-Atoms), the 
Munich Quantum Valley, Berlin Quantum, the 
Quantum 
Flagship (Millenion, PasQuans2), the European Research Council (DebuQC),
the Clusters of Excellence (MATH+, ML4Q), and the DFG (CRC 183, SPP 2514). This work was furthermore supported by the French National Research Agency (ANR) and the Deutsche Forschungsgemeinschaft (DFG) through the Franco-German collaborative project Bosonic Lattice Codes (BoLaCo), ANR Grant No. ANR-24-CE92-0076 and DFG Project No. 545592371. For BoLaCo, this work is the result of an instrumental joint node collaboration.

\appendix

\section{Random construction for LDLCs}
\label{sec:randLDLC}

The algorithm applied by Sommer \emph{et al.} to generate LDLCs is shown in Algorithm~\ref{alg:ldlc}.
The construction guarantees that $H$ is a $(d, d)$-regular sparse matrix for which each column and row contains each value $\{ h_1, h_2, \dots, h_d \}$ exactly once (up to a sign).
The algorithm for the removal is shown in \algref{alg:loop_removal}.

\begin{algorithm}[!htb]
\caption{Generate LDLC parity check matrix}
\label{alg:ldlc}
\DontPrintSemicolon
\SetAlgoLined
\KwIn{Block length $n$, degree $d$, nonzero elements $\{h_1, h_2, \ldots, h_d\}$.}
\KwOut{Magic square LDLC parity check matrix $H$ with generating sequence $\{h_1, h_2, \ldots, h_d\}$.}
$P \leftarrow$ matrix of size $d \times n$ with zeros.
\For{$k = 1$ \KwTo $d$}{
    Assign a random permutation of length $n$ to the $k$-th row of $P$;
}
P $\leftarrow$ loop\_removal!(P, $d$, $n$) \tcp*{See Alg.~\ref{alg:loop_removal}}
\tcp{Check-matrix construction based on P}
Initialize matrix $H$ of size $n \times n$ with zeros;
\For{$i \leftarrow 1$ \KwTo $n$}{
    \For{$j \leftarrow 1$ \KwTo $d$}{
        $H_{P_{j, i}, i} \leftarrow h_{j} \times$ random value in $\{-1, 1\}$;
    }
}
\Return $H$.

\end{algorithm}

\begin{algorithm}[!htb]
\SetAlgoLined
\KwIn{Permutation matrix $P \in \{1, 2, \dots, n\}^{d \times n}$, integers $d, n$, maximum number of iterations max\_iters}
\KwOut{Boolean indicating whether loop removal was successful}

$c \gets 1$\;
loopless\_cols $\gets 0$\;
iters $\gets 0$\;

\While{loopless\_cols $< n$}{
    changed\_perm $\gets 0$\;
    
    \tcp{Check for 2-loops in column $c$}
    \If{any duplicates in $P[:,c]$}{
        Identify first duplicate row index $i$\;
        changed\_perm $\gets i$\;
    }
    \Else{
        \tcp{Check for 4-loops with other columns}
        \For{each column $c_0 \neq c$}{
            common\_elems $\gets \text{intersection}(P[:,c], P[:,c_0])$\;
            \If{$|\text{common\_elems}| \ge 2$}{
                Find first row $i$ in $c$ matching common\_elems[1]\;
                changed\_perm $\gets i$\;
                \textbf{break}\;
            }
        }
    }
    
    \If{changed\_perm $\neq 0$}{
        $i \gets$ random integer in $[1,n]$\;
        Swap $P[\text{changed\_perm}, c]$ with $P[\text{changed\_perm}, i]$\;
        loopless\_cols $\gets 0$\;
    }
    \Else{
        loopless\_cols $\gets$ loopless\_cols + 1\;
    }
    
    $c \gets c + 1$\;
    \If{$c > n$}{ $c \gets 1$ }
    
    iters $\gets$ iters + 1\;
    \If{iters $= \text{max\_iters}$}{\Return false\;}
}

\Return true\;
\caption{Loop removal from permutation matrix $P$~\cite{ldlc}}
\label{alg:loop_removal}
\end{algorithm}

\begin{algorithm}[!htb]
\caption{Iterative LDLC decoder}
\label{alg:ldlc2e}

\KwIn{Parity-check matrix $H$, received vector $\vect{y}$, noise variance $\sigma^2$, number of iterations $T$}
\KwOut{Estimate $\hat{\vect{x}}$ of transmitted lattice point}

\For{each variable node $v_k$ and each neighboring check node $c_j$}{
    $f_{k\to j}^{(0)}(x) \gets \frac{1}{\sqrt{2\pi\sigma^2}} \exp\Big[-\frac{(y_k - x)^2}{2\sigma^2}\Big]$
}

\For{$t = 1$ \KwTo $T$}{

    \For{each check node $c_j$}{
        \For{each variable node $v_k \in \mathcal{N}(j)$}{
            $p_{j\to k}(x) \gets \bigstar_{i \in \mathcal{N}(j)\setminus\{k\}} f_{i\to j}^{(t-1)}\big(x/H_{ij}\big)$\;
            $p'_{j\to k}(x) \gets p_{j\to k}(-H_{jk} x)$\;
            $g_{j\to k}(x) \gets \sum_{l \in \mathbb{Z}} p'_{j\to k}\big(x - l/H_{jk}\big)$\;
        }
    }

    \For{each variable node $v_k$}{
        \For{each check node $c_j \in \mathcal{N}(k)$}{
            $q_{k\to j}(x) \gets f_k^{(0)}(x) \prod_{i \in \mathcal{N}(k)\setminus\{j\}} g_{i\to k}^{(t)}(x)$\;
            $f_{k\to j}^{(t)}(x) \gets q_{k\to j}(x) / \int_\mathbb{R} q_{k\to j}(z)\, dz$\;
        }
    }
}

\For{each variable node $v_k$}{
    $f_k^F(x) \gets f_k^{(0)}(x) \prod_{i \in \mathcal{N}(k)} g_{i\to k}^{(T)}(x)$\;
    $\left(\tilde{\vect{x}}\right)_k \gets \argmax_x f_k^F(x)$\;
}

$\hat{\vect{a}} \gets \lfloor H \tilde{\vect{x}} \rceil$, $\hat{\vect{x}} \gets G \hat{\vect{a}}$\;
\Return $\hat{\vect{a}}, \hat{\vect{x}}$.
\end{algorithm}

\section{Quantized decoder implementation \label{app:quantized_decoder}}
In this section, we review and discuss implementation details of the quantized LDLC decoder we employ in this work.

We begin by describing the original implementation by Sommer~\emph{et al.}~\cite{ldlc}, often referred to as the ``quantized'' decoder.
Within this implementation, each probability density function (PDF) $f_j(x)$ is approximated by a discrete vector $\vec{f}_j(x)$ representing the evaluation on an interval $x$ with finite resolution $\Delta$ and finite range.
Given that shift errors are sampled from a Gaussian distribution of standard deviation $\sigma$, one can guarantee that the truncation error is exponentially small by choosing a range that is a multiple of $\sigma$.
There is no direct rule for setting the resolution $\Delta$; it should be chosen small enough such that quantization errors become negligible without unnecessarily increasing storage requirements.

The check node update requires a convolution of the incoming messages.
To this end, we first map the incoming messages to the unit interval, that is, 
\begin{equation}
    \overline f_{i\to j}^{(t)}(u)
    =
    \sum_{l\in\mathbb Z}
    \frac{1}{|H_{j,i}|}
    f_{i\to j}^{(t)}
    \left(
        \frac{u+l}{H_{j,i}}
    \right),
\end{equation}
with $u =H_{j,i}x_i\bmod 1$.
Numerically, we circumvent the issue that transformed points generally do not align with the discretized grid we use for message representation by splitting the PDF's value into adjacent grid points.
We further use the fact that in Fourier space, convolution is a simple product.
Hence, we can compute the outgoing message by first taking the fast Fourier transform (FFT), multiplying the result, and then taking the inverse FFT. 
Further, since our messages are now periodic, it is sufficient to take the Fourier transform over $1/\Delta$ grid points instead of the full direct range.
In particular, for a message from $c_j$ to $v_{i_k}$, we have
\begin{equation}
    \overline p_{j\to i_k}
    =
    \mathcal F^{-1}
    \left[
        \prod_{i\in\mathcal N(j)\setminus\{i_k\}}
        \mathcal F
        \left(
            \overline f_{i\to j}^{(t)}
        \right)
    \right],
\end{equation}
from which the outgoing message $g_{j \to i_k}$ is obtained by periodic linear interpolation.

The variable node update is simple and corresponds to a multiplication of the quantized messages. 
Following Sommer~\emph{et al.} we widen messages before multiplication, which reduces errors due to quantization of vanishing variance peaks.
To be explicit, instead of $g_{i_l \to k}^{(t-1)}(x)$ we use
\begin{align}
    \tilde{g}_{i_l \to k}^{(t-1)}(x) = g_{i_l \to k}^{(t-1)}(x - \Delta) + g_{i_l \to k}^{(t-1)}(x) +g_{i_l \to k}^{(t-1)}(x + \Delta).
\end{align}

The final marginal computation then follows directly from the variable node update description.

\section{Gaussian mixture decoder \label{app:gaussian_mixture_decoder}}
In this implementation of the decoder, often referred to as Gaussian mixture decoder~\cite{kurkoski_message-passing_2008}, messages are represented through Gaussian components described by their amplitude within the mixture, as well as their mean and variance.
Since the number of components would grow exponentially in the number of iterations, components need to be truncated throughout the algorithm.

Various improvements upon the original description in Ref.~\cite{kurkoski_message-passing_2008} followed in the classical literature on low-density lattice codes~\cite{kurkoski_single-gaussian_2009, wang_efficient_2023, liu_efficient_2019} which differ mostly in the number of Gaussian components kept within each message, and thus, consequently, how variable node updates are performed.

In contrast to the quantized decoder discussed above, the check node update is the simpler update rule in this representation.
In particular, check node $c_j$ sends a message $g_{j\to i_k}$ to each neighboring variable node $v_{i_k}$ which is given by~\cite{kurkoski_message-passing_2008}
\begin{align}
    g_{j\to i_k} = \sum_{i = -\infty}^{\infty} \mathcal{N}(x, m_{k} - i / H_{j, i}, \sigma_k^2),
\end{align}
with 
\begin{align}
    m_k = - \frac{1}{H_{j, k}} \sum_{i \neq k} H_{j, i} m_{v, i}, \quad \sigma^2_k = \frac{1}{H_{j,k}^2} \sum_{i \neq k} H_{j, i}^2 \sigma_{v, i}^2.
\end{align}

Below, we describe in detail, two approaches towards realizing the variable node update that we refer to as \emph{nearest} and \emph{list sphere decoding} (\emph{LSD}) that are originally due to Ref.~\cite{liu_efficient_2019} and Ref.~\cite{wang_efficient_2023}, respectively.
In~\appref{app:var_node_memory} we also describe how to add memory to the variable node update.

\subsection{Variable node updates}

\label{sec:LDLCBP}
In this section, we discuss several strategies for computing the variable-to-check messages, which involve approximating the product of infinite trains of Gaussian functions by a finite sum. 
The variable-node update constitutes one of the most computationally intensive steps in LDLC decoding. 
Although various approximations have been proposed to reduce this cost, they often come at the expense of decoding accuracy or robustness.

\subsubsection{The list-sphere decoding subroutine \label{ssec:lsd_routine}}
Here, we rederive an efficient version of the \emph{list-sphere decoding} (LSD) variable-node update routine originally proposed in Ref.~\cite{wang_efficient_2023}, refining the derivation and correcting a few inconsistencies to provide a clear and self-contained formulation.

In the variable node message, the following product must be computed at each node,
\begin{align}
    \tilde{f}_k(x) & =\mathcal{N}\left(x ; y_k, \sigma_k^2\right) \prod_{l=1}^{d-1} p_l(x) \\
    & =\mathcal{N}\left(x ; y, \sigma^2\right) \prod_{l=1}^{d-1} \sum_{z_l \in \mathcal{Z}} \mathcal{N}\left(x ; m_l+\frac{z_l}{h_l}, \sigma_l^2\right),
    \nonumber
\end{align}
where $y_k$ and $\sigma_k^2$ are the mean and variance of the local message at variable node $v_k$, 
$m_l$ and $\sigma_l^2$ denote the mean and variance of the $l$-th incoming check-to-variable message, 
and $h_l = H_{k,l}$ is the corresponding entry of the parity-check matrix.

Focusing on a single combination of integer shifts $\{z_l\}$, the product of Gaussians can be written as
\begin{align}
    c \mathcal{N}(x, m, \Delta^2) = \mathcal{N}(x; y_k, \sigma_k^2) \prod_{l=1}^{d - 1} \mathcal{N}(x, m_l + \frac{z_l}{h_l}, \sigma_l^2).
\end{align}
Note that we can absorb the local variable message into the product term by denoting $\sigma_d^2 = \sigma_k^2$, $m_d = y_k$, and $h_d = 1$. 
By doing so, we can 
express the variance $\Delta^2$ and the mean $m$ of the product through the Gaussian product rule~\cite{ldlc}, that is,
\begin{align}
    \Delta^{-2} &= \sum_{l = 1}^{d} \frac{1}{\sigma_l^2},\\
    m &= \Delta^2 \sum_{l = 1}^{d} \frac{m_l + z_l / h_l}{\sigma_l^2},
\end{align}
and the scaling 
factor 
$c$ is given by
\begin{align}
    \label{app-eq:gaussian_scaling_factor}
    c =& \frac{1}{\sqrt{(2\pi)^{d-1} \Delta^{-2} \prod_{l=1}^{d} \sigma_l^2}} 
        \\ &
        \times \exp\Bigg[-\frac{\Delta^2}{2} \sum_{i=1}^{d-1} \sum_{j=i+1}^{d} 
        \frac{(m_i + z_i / h_i - m_j - z_j / h_j)^2}{\sigma_i^2 \sigma_j^2}\Bigg].
        \nonumber
\end{align}
The pre-factor depends only on the variances $\sigma_l^2$ and not on the integer shifts $z_l$, so it can be treated as a constant.  
Moreover, the exponent has a quadratic form in the $z_l$’s, which we exploit in the following by writing
\begin{align}
    c \propto \exp(- \frac{1}{2} (\vec{z} + \vec{p})^{T} \mat{Q} (\vec{z} + \vec{p})),
\end{align}
where $\vec{p} = \vec{h} \odot \vec{m}$ and the matrix $\mat{Q}$
\begin{align}
    \mat{Q} &= \text{diagm}(\frac{1}{h_1^2 \sigma_1^2}, \dots, \frac{1}{h_d^2 \sigma_d^2}) - \vec{q}^T \vec{q}, \\
    \vec{q} &= (\frac{\Delta}{h_1 \sigma_1^2}, \dots, \frac{\Delta}{h_d \sigma_d^2}).
\end{align}
To approximate the variable node message, we seek the set of integer vectors $\vec{z}$ that satisfy
\begin{align}
    0 < (\vec{z} + \vec{p})^T \mat{Q} (\vec{z} + \vec{p}) < \beta^2,
\end{align}
where $\beta \in \mathbb{R}$ is a parameter controlling the \emph{quality} of the approximation.

By performing a Cholesky decomposition of $\mat{Q}$, i.e., $\mat{Q} = \mat{R}^T \mat{R}$, this problem can be reformulated as a closest vector search in a $(d-1)$-dimensional lattice
\begin{align}
    ||\mat{R} (\vec{z} + \vec{p})^T|| < \beta^2.
    \label{eq:lsd_closest_vector_problem}
\end{align}

While this formulation is conceptually straightforward, computing the full Cholesky factorization generally requires $\mathcal{O}(d^3)$ operations, which can be prohibitive for large $d$. Fortunately, as observed in Ref.~\cite{wang_efficient_2023}, only the diagonal elements of $\mat{R}$ are necessary to identify the relevant integer vectors, allowing the computation to be reduced to $\mathcal{O}(d)$ time.

To simplify the computation, we first rewrite $\mat{Q}$ in a scaled form as
\begin{align}
   \mat{S}^{-1} \mat{Q} \mat{S}^{-1}
   = \mat{S}^{-1} \mat{R}^{T} \mat{R} \mat{S}^{-1}
   = \tilde{\mat{R}}^{T} \tilde{\mat{R}}
   = \mathbbm{1} - \vec{t}^T \vec{t},
\end{align}
where
\begin{align}
    \vec{t} &= \left(\sign(h_1)\frac{1}{\sigma_1}, \dots, \sign(h_d)\frac{1}{\sigma_d}\right), \\
    \mat{S} &= \operatorname{diag}\!\left(\frac{1}{\sqrt{h_1^2 \sigma_1^2}}, \dots, \frac{1}{\sqrt{h_d^2 \sigma_d^2}}\right),
\end{align}
and the two Cholesky factors are related through $\mat{R} = \tilde{\mat{R}} \mat{S}$.

For a matrix of the form $\mathbbm{1} - \vec{t}^T \vec{t}$, an explicit Cholesky factorization $\tilde{\mat{R}}$ has been derived in Ref.~\cite{wen_efficient_2016}. Using that result, the elements of $\mat{R}$ can be written as
\begin{align}
    R_{ij} =
    \begin{cases}
        \displaystyle
        \frac{1}{\sqrt{h_j^2 \sigma_j^2}}
        \sqrt{\frac{1 - \sum_{l=1}^{i} t_l^2}{1 - \sum_{l=1}^{i-1} t_l^2}},
        & j = i, \\[1.2em]
        \displaystyle
        \frac{1}{\sqrt{h_j^2 \sigma_j^2}}
        \frac{-t_i t_j}{
        \sqrt{1 - \sum_{l=1}^{i} t_l^2}
        \sqrt{1 - \sum_{l=1}^{i-1} t_l^2}},
        & i < j.
    \end{cases}
\end{align}

Next, we expand Eq.~\eqref{eq:lsd_closest_vector_problem} using the upper-triangular structure of $\mat{R}$, to get
\begin{align}
    \|\mat{R}(\vec{z} + \vec{p})^T\|^2 
    = \sum_{i=1}^{d} R_{i,i}^2 \left( z_i + p_i + \sum_{j=i+1}^{d}  \frac{R_{i,j} (z_j + p_j)}{R_{i,i}} \right)^2. 
    \label{eq:cvp_rewrite}
\end{align}
In order to simplify this expression, we introduce
\begin{align}
    \gamma_i \coloneqq  -p_i - \frac{1}{R_{i,i}} \sum_{j=i+1}^{d} R_{i,j} (z_j + p_j),
\end{align}
so that $\gamma_d = -p_d$. With this definition, Eq.~\eqref{eq:cvp_rewrite} becomes
\begin{align}
    \sum_{i=1}^{d} R_{i,i}^2 (z_i - \gamma_i)^2 < \beta^2.
\end{align}
Substituting the explicit expression for $R_{i,j}/R_{i,i}$,
we find
\begin{align}
    \frac{R_{i,j}}{R_{i,i}} = - \frac{t_i t_j}{1 - \sum_{l=1}^{i} t_l^2} \sqrt{\frac{h_i^2 \sigma_i^2}{h_j^2 \sigma_j^2}},
\end{align}
and thus obtain
\begin{align}
    \gamma_i = -p_i + \sum_{j=i+1}^{d} \sqrt{\frac{h_i^2 \sigma_i^2}{h_j^2 \sigma_j^2}} \frac{t_i t_j}{1 - \sum_{l=1}^{i} t_l^2} (z_j + p_j).
\end{align}

This expression can be further recast into an iterative form, solved starting from $i = d-1$, and iterated as
\begin{align}
    \gamma_i = -p_i + \frac{g_i t_i}{f_i} u_{i+1},
\end{align}
where
\begin{align*}
    g_i &= \sqrt{h_i^2 \sigma_i^2}, & f_i &= 1 - \sum_{l=1}^{i} t_l^2, & u_i &= \frac{t_i (z_i + p_i)}{g_i} + u_{i+1}.
\end{align*}
The recursion is initialized with $u_{d+1} = 0$, and from the boundary condition $\gamma_d = -p_d$, one finds $u_d = y_k/\sigma_d^2$. 
With this iterative formulation, we are now in a position to present the variable node update algorithm (Algorithm~\ref{algo:lsd_update}), which is a modified version of the method in Ref.~\cite{agrell_closest_2002}.

\begin{algorithm}[!htb]
\caption{Efficient variable node list sphere decoding}
\label{algo:lsd_update}
\SetKwInOut{Input}{Input}
\SetKwInOut{Output}{Output}

\Input{$\vec{f}, \vec{g}, \vec{p}, \vec{R_{i, i}}, \beta$}
\Output{$\mathcal{L}$}

$d \gets$ length($\vec{p}$) \tcp*{Problem size}
$\mathcal{L} \gets \emptyset$ \,;
$\vec{z} \gets \mathbf{0}$ \tcp*{Solution vector}
$\vec{dist} \gets \mathbf{0}$ \,;
$\vec{u} \gets \mathbf{0}$ \,;
$k \gets d - 1$ \tcp*{Start from second-to-last index}
$u_d \gets y_k / \sigma_d^2$ \tcp*{Initial condition}
$\gamma_k \gets -p_k + \frac{t_k g_k}{f_k} \cdot u_{k+1}$ \,;
$z_k \gets \text{round}(\gamma_k)$ \,;
$s_k \gets \text{sgn}(\gamma_k - z_k)$ \,;
$dist_k \gets dist_{k+1} + R_{kk}^2 (z_k - \gamma_k)^2$ \,;

\While{$k \le d-1$}{
    \eIf{$dist_k < \beta^2$}{
        \eIf{$k == 1$}{
            Add $[\vec{z}, 0, dist]$ to $\mathcal{L}$\,;
            $z_k \gets z_k + s_k$\,;
            $s_k \gets -s_k - \text{sgn}(s_k)$\,;
            $dist_k \gets dist_{k+1} + R_{kk}^2 (z_k - \gamma_k)^2$\,;
        }{
            $u_k \gets \frac{t_k (z_k + p_k)}{g_k} + u_{k+1}$\,;
            $k \gets k - 1$\,;
            $\gamma_k \gets -p_k + \frac{t_k g_k}{f_k} \cdot u_{k+1}$\,;
            $z_k \gets \text{round}(\gamma_k)$\,;
            $s_k \gets \text{sgn}(\gamma_k - z_k)$\,;
            $dist_k \gets dist_{k+1} + R_{k,k}^2 (z_k - \gamma_k)^2$\,;
        }
    }{
        \eIf{$k == d-1$}{
            \Return $\mathcal{L}$\,;
        }{
            $k \gets k + 1$\,;
            $z_k \gets z_k + s_k$\,;
            $s_k \gets -s_k - \text{sgn}(s_k)$\,;
            $dist_k \gets dist_{k+1} + R_{k,k}^2 (z_k - \gamma_k)^2$\,;
        }
    }
}
\end{algorithm}

\subsubsection{Two-Gaussian approximation for variable node updates}
In the \emph{Gaussian-approximation} (GA) LDLC decoding \cite{liu_efficient_2019}, the variable-to-check messages
$\hat f_j(w)$ are approximated as a mixture of only two Gaussians per incoming edge, instead of the full infinite periodic sum
\begin{equation}
\hat f_j(w) \approx \mathcal{N}(w; y_k, \sigma^2_k) \prod_{i=1, i \neq j}^d \bigl( \mathcal{N}_{L,i} + \mathcal{N}_{R,i} \bigr),
\end{equation}
where $\mathcal{N}_{L,i}$ and $\mathcal{N}_{R,i}$ are the two Gaussian components of the $i$-th check message
closest to the current variable node mean $y_k$.  

This approximation is justified because in the product of two Gaussian functions with means $m_1$ and $m_2$, the scaling factor $c$ (see Eq.~\eqref{app-eq:gaussian_scaling_factor}) exponentially approaches zero if $(m_1 - m_2)^2 / (\sigma_1^2 + \sigma_2^2) \gg 1$.
The pseudocode for the corresponding variable node update rules is given in \algref{algo:ga_ldlc}, which utilizes the \textsc{nearest} subroutine described in \algref{algo:nearest}.

\begin{algorithm}[!htb]
\caption{Two-Gaussian approximation: Variable node update} \label{algo:ga_ldlc}
\SetAlgoLined
\KwIn{Incoming check messages $\{\mathcal{N}_i\}_{i=1}^{d}$, variable node mean $y_k$, variance $\sigma^2_k$, selection radius $\epsilon_r$}.
\KwOut{Outgoing check messages $\{f_j(w)\}_{j=1}^{d}$}.

\For{$i = 1$ \KwTo $d$}{   
    $\mathcal{N}_{L, i}, \mathcal{N}_{R, i} \gets \textsc{nearest}(\mathcal{N}_i, y_k, H_{i,k}, \epsilon_r)$;
}

\For{$j = 1$ \KwTo $d$}{
    $c_L \mathcal{N}_L \gets \mathcal{N}(y_k, \sigma_k^2) \prod_{i=1, i \neq j} \mathcal{N}_{L, i}$;
    $c_R \mathcal{N}_R \gets \mathcal{N}(y_k, \sigma_k^2) \prod_{i=1, i \neq j} \mathcal{N}_{R, i}$;

    $m_{v,j} \gets \frac{c_L m_L + c_R m_R}{c_L + c_R}$;
    $\sigma_{v,j}^2 \gets \frac{c_L (\sigma_L^2 + m_L^2) + c_R (\sigma_R^2 + m_R^2)}{c_L + c_R} - m_{v,j}^2$;

    $f_j(w) \gets \mathcal{N}(w; m_{v,j}, \sigma_{v,j}^2)$;
}

\Return $\{f_j(w)\}_{j=1}^{d}$.
\end{algorithm}

\begin{algorithm}[!htb]
\caption{\textsc{nearest} subroutine} \label{algo:nearest}
\SetAlgoLined
\KwIn{Gaussian message $\mathcal{N}(w; m, \sigma^2)$, variable node mean $y_k$, weight $H_{jk}$, selection radius $\epsilon_r$}
\KwOut{Nearest Gaussian pair $\mathcal{N}_L, \mathcal{N}_R$}

$a \gets \lfloor (m - y_k) / H_{jk} \rfloor$.
$b \gets a + 1$.

$m_L \gets m - a / H_{j,k}$.
$m_R \gets m - b / H_{j,k}$.

\If{$(y_k - \epsilon_r) < m_L < (y_k + \epsilon_r)$ \textbf{and} $\neg((y_k - \epsilon_r) < m_R < (y_k + \epsilon_r))$}{
    $m_R \gets m_L$.
}
\ElseIf{$(y_k - \epsilon_r) < m_R < (y_k + \epsilon_r)$ \textbf{and} $\neg((y_k - \epsilon_r) < m_L < (y_k + \epsilon_r))$}{
    $m_L \gets m_R$.
}

\If{$m_L > m_R$}{
    $m_L, m_R \gets m_R, m_L$.
}

\Return $\mathcal{N}(m_L, \sigma^2), \mathcal{N}(m_R, \sigma^2)$.
\end{algorithm}

\subsection{Variable-node memory \label{app:var_node_memory}}
Quite generally, we can add memory to all variable-node updates as is common in the (classical) discrete-variable literature.
In particular, we introduce memory in the following way.
Let \begin{equation}
q_i(x)=\mathcal{N}\!\left(x;y_i,\sigma_i^2\right)
\end{equation}
denote the original channel message at variable node $i$, and let
\begin{equation}
p_i^{(t)}(x) = \mathcal{N}\!\left(x;\mu_i^{(t)},v_i^{(t)}\right)
\end{equation}
denote the Gaussian approximation to the variable-node posterior at iteration
$t$, obtained by combining the channel information with all incoming
check-node messages. 
We then introduce memory, also known as momentum, by forming the two-component Gaussian mixture
\begin{align}
\widetilde{q}_i^{(t)}(x) = (1-\gamma_i)q_i(x) + \gamma_i p_i^{(t)}(x),
\end{align}
where $\gamma$ is the memory strength. 
Using moment-matching as described above, we reduce the mixture into a single Gaussian, which is used in the following iteration of the algorithm.

\section{Dimensionality reduction of GKP-LDLC codes}\label{app:dim_red}

Algorithm~\ref{alg:dim_red} summarizes the single-qubit reduction of \secref{sec:dim_red}. Starting from a trivial GKP-LDLC generator $M=\sqrt{d}H$, it computes the symplectic Gram matrix $A=MJM^T$ and its integer symplectic canonical form $F=CAC^T$ (with $C$ unimodular), which exposes the invariant factors $d_1\mid\cdots\mid d_n$. After reduction, the code contains a single logical qubit exactly when $d_1=\cdots=d_{n-1}=1$ and $d_n=2\ell$ is even. Concretely, the reduction rescales the two dense canonical rows carrying the factor $2\ell$ by a balanced factorization $\ell=k_1k_2$, producing a determinant-$2$ generator with logical operators $X_L=\tilde{\vect u}/2$, $Z_L=\tilde{\vect v}/2$ and $Y_L=X_L+Z_L$.

All algebraic steps (the canonical form, the factor splitting, and the reconstruction of the logical coset representatives) can be carried out in exact rational arithmetic (with arbitrary-precision integers), so that the returned generator provably satisfies $\lvert\det\tilde{M}\rvert=2$ together with the logical (anti)commutation relations. The only floating-point step in our implementation is the closest-vector search used to find short logical representatives. Because the raw generator is extremely skewed, it is first reduced \emph{exactly} over the integers (an exact LLL reduction after clearing denominators), and the CVP solver is then used only to pick \emph{integer} lattice coordinates, which are lifted back to exact rationals. Once the lattice coordinates have been selected, the corresponding representatives and their norms are evaluated exactly.

\section{Optimizing local dimensions via invariant factors}\label{app:invariants}

The reduction of \secref{sec:dim_red} applies verbatim only when the invariant factors of the trivial GKP-LDLC code have the special form $(1,\dots,1,2\ell)$. More generally, the canonical form $F=CAC^T=J_2\otimes D$ has a diagonal $D=\mathrm{diag}(d_1,\dots,d_n)$ of invariant factors $d_1\mid d_2\mid\cdots\mid d_n$, and the $j$-th mode of the canonical code carries logical dimension $d_j$. The largest factor $\max_j d_j$ is the maximum local dimension one has to contend with; the smaller it is, the closer the code is to a genuine qubit code and, empirically, the larger its normal-form distance.

Two GKP codes are equivalent under a Gaussian unitary if and only if their integer symplectic Gram matrices are congruent, $A'=UAU^T$ for a unimodular $U$~\cite{Burchards2025fiberbundlefault}. The congruence class of an antisymmetric integer form is completely characterized by its \emph{Pfaffian divisors}
\begin{equation}
    \pi_k = \gcd\!\Big( \textstyle\prod_{j\in S} d_j \;:\; S\subseteq\{1,\dots,n\},\ |S|=k \Big),
\end{equation}
with $k=1,\dots,n$, or, equivalently by the divisibility-ordered invariant factors $s_1\mid\cdots\mid s_n$ (with $\pi_k=s_1\cdots s_k$). 
The canonical form $J_2\otimes\mathrm{diag}(s_1,\dots,s_n)$ is unique, but the \emph{same} code---the same congruence class---can also be realized by a block-diagonal form $J_2\otimes\mathrm{diag}(d'_1,\dots,d'_n)$ whose entries need \emph{not} obey the divisibility chain, provided they reproduce the same Pfaffian divisors. Since the $j$-th mode of such a realization carries local dimension $d'_j$, we are free to choose, among all congruent diagonals, the one that minimizes $\max_j d'_j$.

The mechanism is clearest for two modes. Consider a code whose canonical diagonal is $D=\mathrm{diag}(1,36)$, i.e.\ a single mode carrying local dimension $36=2^2\cdot 3^2$; its Pfaffian divisors are $\pi_1=\gcd(1,36)=1$ and $\pi_2=1\cdot 36=36$. The diagonal $D'=\mathrm{diag}(4,9)$ has exactly the same Pfaffian divisors ($\pi_1=\gcd(4,9)=1$, $\pi_2=4\cdot 9=36$) and hence realizes a code in the same Gaussian-unitary equivalence class, yet its maximum local dimension is only $9$ instead of $36$: the two prime powers $2^2$ and $3^2$ have been placed on different modes rather than piled onto one. The improvement requires at least two distinct primes; a single prime power cannot be redistributed while preserving the Pfaffian divisors---for instance $\mathrm{diag}(1,1,8)$ with $8=2^3$ admits no rebalancing, and its maximum local dimension stays $8$.

Concretely, this optimization factorizes the invariant factors and, for each prime independently, permutes its valuations across the modes---anti-correlating distinct primes---so as to minimize $\max_j d'_j$, checking at each step that the Pfaffian divisors are unchanged. Applied to the trivial GKP-LDLC codes, it lowers the largest mode-wise logical dimension and, correspondingly, tends to raise the normal-form distance before the single-qubit reduction of \algref{alg:dim_red} is attempted. This is the invariant-based local-dimension optimization implemented in the \textsc{SymplecticGKP.jl} package~\cite{julia_symplectic_gkp}.

\bibliography{references}

@misc{julia_lattice_decoder,
  author = {Timo Hillmann and Francesco Arzani},
  title    = {Lattice{D}ecoder.jl},
  version  = {0.1.0},
  license  = {MIT},
  url      = {https://github.com/timohillmann/LatticeDecoder.jl},
  publisher = {GitHub},
  year     = {2026},
  month    = sep
}

@misc{julia_symplectic_gkp,
  author = {Francesco Arzani and Timo Hillmann},
  title    = {Symplectic{GKP}.jl},
  version  = {0.1.0},
  license  = {GPL-3.0},
  url      = {https://github.com/frarzani/SymplecticGKP.jl},
  publisher = {GitHub},
  year     = {2026},
  month    = sep
}

@inproceedings{aggarwal2015solvingclosestvectorproblem,
  author    = {Aggarwal, Divesh and Dadush, Daniel and
               Stephens-Davidowitz, Noah},
  title     = {Solving the {closest vector problem} in {$2^n$}
               time---The {discrete Gaussian} strikes again!},
  booktitle = {2015 IEEE 56th Annual Symposium on Foundations of
               Computer Science (FOCS)},
  pages     = {563--582},
  year      = {2015},
  eprint = {1504.01995},
  optprimaryclass = {cs.DS},
  archiveprefix = {arXiv},
  publisher = {IEEE},
  doi       = {10.1109/FOCS.2015.41},
  url       = {https://doi.org/10.1109/FOCS.2015.41}
}

@article{aghaee_rad_scaling_2025,
  title = {Scaling and Networking a Modular Photonic Quantum Computer},
  shorttitle = {Xanadu Bosonic},
  author = {Aghaee Rad, H. and {others}},
  year = 2025,
  month = feb,
  journal = {Nature},
  volume = {638},
  number = {8052},
  pages = {912--919},
  issn = {1476-4687},
  doi = {10.1038/s41586-024-08406-9},
  url = {https://www.nature.com/articles/s41586-024-08406-9},
  urldate = {2025-03-01},
  copyright = {2025 The Author(s)},
  langid = {english}
}

@article{agrell_closest_2002,
  title = {Closest Point Search in Lattices},
  author = {Agrell, E. and Eriksson, T. and Vardy, A. and Zeger, K.},
  year = 2002,
  month = aug,
  journal = {IEEE Trans. Inf. Th.},
  volume = {48},
  number = {8},
  pages = {2201--2214},
  issn = {1557-9654},
  doi = {10.1109/TIT.2002.800499}
}

@inproceedings{Ajtai,
  title = {Sampling short lattice vectors and the closest lattice vector problem},
  booktitle = {Proceedings of the 17th Annual IEEE Conference on Computational Complexity (CCC 2002)},
  author = {Ajtai, Mikl{\'o}s and Kumar, Ravi and Sivakumar, D.},
  year = {2002},
  pages = {53--57},
  publisher = {IEEE Computer Society},
  doi = {10.1109/CCC.2002.1004339},
  url = {https://doi.org/10.1109/CCC.2002.1004339}
}

@inproceedings{arora_hardness_1993,
  title = {The Hardness of Approximate Optima in Lattices, Codes, and Systems of Linear Equations},
  booktitle = {Proceedings of 1993 {{IEEE}} 34th {{Annual Foundations}} of {{Computer Science}}},
  author = {Arora, S. and Babai, L. and Stern, J. and Sweedyk, Z.},
  year = 1993,
  month = nov,
  pages = {724--733},
  doi = {10.1109/SFCS.1993.366815},
  url = {https://ieeexplore.ieee.org/document/366815},
  urldate = {2025-10-23}
}

@article{babai_lovasz_1986,
  title = {On {{Lov\'asz}}' Lattice Reduction and the Nearest Lattice Point Problem},
  author = {Babai, L.},
  year = 1986,
  month = mar,
  journal = {Combinatorica},
  volume = {6},
  number = {1},
  pages = {1--13},
  issn = {1439-6912},
  doi = {10.1007/BF02579403},
  url = {https://doi.org/10.1007/BF02579403},
  urldate = {2025-11-05},
  langid = {english}
}

@article{Burchards2025fiberbundlefault,
  doi = {10.22331/q-2025-10-29-1899},
  url = {https://doi.org/10.22331/q-2025-10-29-1899},
  title = {Fiber {b}undle {f}ault {t}olerance of {GKP} {c}odes},
  author = {Burchards, Ansgar G. and Flammia, Steven T. and Conrad, Jonathan},
  journal = {{Quantum}},
  issn = {2521-327X},
  publisher = {{Verein zur F{\"{o}}rderung des Open Access Publizierens in den Quantenwissenschaften}},
  volume = {9},
  pages = {1899},
  month = oct,
  year = {2025},
  eprint = {2410.07332},
  optprimaryclass = {quant-ph},
  archiveprefix = {arXiv},
}

@article{Conrad2024goodgottesmankitaev,
  doi = {10.22331/q-2024-07-04-1398},
  url = {https://doi.org/10.22331/q-2024-07-04-1398},
  title = {Good {G}ottesman-{K}itaev-{P}reskill codes from the {NTRU} cryptosystem},
  author = {Conrad, Jonathan and Eisert, Jens and Seifert, Jean-Pierre},
  journal = {{Quantum}},
  publisher = {{Verein zur F{\"{o}}rderung des Open Access Publizierens in den Quantenwissenschaften}},
  volume = {8},
  pages = {1398},
  month = jul,
  year = {2024},
  eprint = {2303.02432},
  optprimaryclass = {quant-ph},
  archiveprefix = {arXiv},
}

@article{Conrad_2022,
  ids = {conrad_gottesman-kitaev-preskill_2021},
  title = {Gottesman-{{Kitaev-Preskill}} Codes: {{A}} Lattice Perspective},
  shorttitle = {Gottesman-{{Kitaev-Preskill}} Codes},
  author = {Conrad, Jonathan and Eisert, Jens and Arzani, Francesco},
  year = 2022,
  month = feb,
  journal = {Quantum},
  volume = {6},
  eprint = {2109.14645},
  pages = {648},
  publisher = {Verein zur F\"orderung des Open Access Publizierens in den Quantenwissenschaften},
  doi = {10.22331/q-2022-02-10-648},
  url = {https://quantum-journal.org/papers/q-2022-02-10-648/},
  urldate = {2022-04-11},
  archiveprefix = {arXiv},
  langid = {british}
}

@phdthesis{conradPhD,
  title = {The fabulous world of {{GKP}} codes},
  author = {Conrad, Jonathan},
  year = 2024,
  school = {Freie Universit{\"a}t Berlin},
  eprint = {2412.02442},
  optprimaryclass = {quant-ph},
  pages = {XII, 149 Seiten},
  doi = {10.17169/refubium-45505},
  url = {http://arxiv.org/abs/2412.02442},
  urldate = {2025-03-20},
  archiveprefix = {arXiv}
}

@book{ConwaySloane:1999,
  title = {Sphere {{packings}}, {{lattices}} and {{groups}}},
  author = {Conway, J. H. and Sloane, N. J. A.},
  editor = {Chern, S. S. and Eckmann, B. and {de la Harpe}, P. and Hironaka, H. and Hirzebruch, F. and Hitchin, N. and H{\"o}rmander, L. and Knus, M.-A. and Kupiainen, A. and Lannes, J. and Lebeau, G. and Ratner, M. and Serre, D. and Sinai, {\relax Ya}. G. and Sloane, N. J. A. and Tits, J. and Waldschmidt, M. and Watanabe, S. and Berger, M. and Coates, J. and Varadhan, S. R. S.},
  year = 1999,
  series = {Grundlehren Der Mathematischen {{Wissenschaften}}},
  volume = {290},
  publisher = {Springer},
  address = {New York, NY},
  doi = {10.1007/978-1-4757-6568-7},
  url = {http://link.springer.com/10.1007/978-1-4757-6568-7},
  urldate = {2023-02-21},
  isbn = {978-1-4757-6568-7}
}

@article{dinur_approximating_2003,
  title = {Approximating {CVP} to within almost-polynomial factors is {NP}-hard},
  author = {Dinur, I. and Kindler, G. and Raz, R. and Safra, S.},
  year = 2003,
  month = apr,
  journal = {Combinatorica},
  volume = {23},
  number = {2},
  pages = {205--243},
  issn = {1439-6912},
  doi = {10.1007/s00493-003-0019-y},
  url = {https://doi.org/10.1007/s00493-003-0019-y},
  urldate = {2025-10-23},
  langid = {english}
}

@misc{eisert2025mindgapsfraughtroad,
  title = {Mind the gaps: The fraught road to quantum advantage},
  author = {Eisert, Jens and Preskill, John},
  year = {2025},
  eprint = {2510.19928},
  optprimaryclass = {quant-ph},
  doi = {10.48550/arXiv.2510.19928},
  url = {https://arxiv.org/abs/2510.19928},
  archiveprefix = {arXiv}
}

@article{fossorier_soft-decision_1995,
  title = {Soft-Decision Decoding of Linear Block Codes Based on Ordered Statistics},
  author = {Fossorier, M.P.C. and Lin, Shu},
  year = 1995,
  month = sep,
  journal = {IEEE Trans. Inf. Th.},
  volume = {41},
  number = {5},
  pages = {1379--1396},
  issn = {1557-9654},
  doi = {10.1109/18.412683}
}

@phdthesis{gallager_low_1960,
  type = {Thesis},
  title = {Low density parity check codes},
  author = {Gallager, Robert G.},
  year = 1960,
  url = {https://dspace.mit.edu/handle/1721.1/11804},
  urldate = {2025-04-14},
  copyright = {M.I.T. theses are protected by copyright. They may be viewed from this source for any purpose, but reproduction or distribution in any format is prohibited without written permission. See provided URL for inquiries about permission.},
  langid = {english},
  school = {Massachusetts Institute of Technology}
}

@article{GKP,
  title = {Encoding a Qubit in an Oscillator},
  author = {Gottesman, Daniel and Kitaev, Alexei and Preskill, John},
  year = 2001,
  eprint = {quant-ph/0008040},
  optprimaryclass = {quant-ph},
  archiveprefix = {arXiv},
  month = jun,
  journal = {Phys. Rev. A},
  volume = {64},
  number = {1},
  pages = {012310},
  publisher = {American Physical Society},
  issn = {1050-2947},
  doi = {10.1103/PhysRevA.64.012310},
  url = {https://link.aps.org/doi/10.1103/PhysRevA.64.012310},
  urldate = {2019-11-22},
  langid = {english}
}

@article{GKPBlueprint,
  title = {Blueprint for a {{scalable photonic fault-tolerant quantum computer}}},
  author = {Bourassa, J. Eli and Alexander, Rafael N. and Vasmer, Michael and Patil, Ashlesha and Tzitrin, Ilan and Matsuura, Takaya and Su, Daiqin and Baragiola, Ben Q. and Guha, Saikat and Dauphinais, Guillaume and Sabapathy, Krishna K. and Menicucci, Nicolas C. and Dhand, Ish},
  year = 2021,
  month = feb,
  journal = {Quantum},
  volume = {5},
  eprint = {2010.02905},
  pages = {392},
  publisher = {Verein zur F\"orderung des Open Access Publizierens in den Quantenwissenschaften},
  doi = {10.22331/q-2021-02-04-392},
  url = {https://quantum-journal.org/papers/q-2021-02-04-392/},
  urldate = {2021-02-26},
  archiveprefix = {arXiv},
  langid = {british}
}

@article{GKPIons,
  title = {Encoding a Qubit in a Trapped-Ion Mechanical Oscillator},
  author = {Fl{\"u}hmann, C. and Nguyen, T. L. and Marinelli, M. and Negnevitsky, V. and Mehta, K. and Home, J. P.},
  year = 2019,
  eprint = {1807.01033},
  optprimaryclass = {quant-ph},
  archiveprefix = {arXiv},
  month = feb,
  journal = {Nature},
  volume = {566},
  number = {7745},
  pages = {513--517},
  publisher = {Nature Publishing Group},
  issn = {1476-4687},
  doi = {10.1038/s41586-019-0960-6},
  url = {https://www.nature.com/articles/s41586-019-0960-6},
  urldate = {2020-07-29},
  copyright = {2019 The Author(s), under exclusive licence to Springer Nature Limited}
}

@article{GKPLight,
  ids = {konno_propagating_2023},
  title = {Logical States for Fault-Tolerant Quantum Computation with Propagating Light},
  author = {Konno, Shunya and Asavanant, Warit and Hanamura, Fumiya and Nagayoshi, Hironari and Fukui, Kosuke and Sakaguchi, Atsushi and Ide, Ryuhoh and China, Fumihiro and Yabuno, Masahiro and Miki, Shigehito and Terai, Hirotaka and Takase, Kan and Endo, Mamoru and Marek, Petr and Filip, Radim and {van Loock}, Peter and Furusawa, Akira},
  year = 2024,
  month = jan,
  journal = {Science},
  volume = {383},
  number = {6680},
  eprint = {2309.02306},
  pages = {289--293},
  doi = {10.1126/science.adk7560},
  url = {https://www.science.org/doi/10.1126/science.adk7560},
  urldate = {2025-03-26},
  archiveprefix = {arXiv}
}

@article{GKPSuperconducting,
  title = {Quantum Error Correction of a Qubit Encoded in Grid States of an Oscillator},
  author = {{Campagne-Ibarcq}, P. and Eickbusch, A. and Touzard, S. and {Zalys-Geller}, E. and Frattini, N. E. and Sivak, V. V. and Reinhold, P. and Puri, S. and Shankar, S. and Schoelkopf, R. J. and Frunzio, L. and Mirrahimi, M. and Devoret, M. H.},
  year = 2020,
  month = aug,
  journal = {Nature},
  volume = {584},
  number = {7821},
  eprint = {1907.12487},
  pages = {368--372},
  publisher = {Nature Publishing Group},
  issn = {1476-4687},
  doi = {10.1038/s41586-020-2603-3},
  url = {https://www.nature.com/articles/s41586-020-2603-3},
  urldate = {2020-10-28},
  archiveprefix = {arXiv},
  copyright = {2020 The Author(s), under exclusive licence to Springer Nature Limited},
  langid = {english}
}

@article{goldreich_approximating_1999,
  title = {Approximating Shortest Lattice Vectors Is Not Harder than Approximating Closest Lattice Vectors},
  author = {Goldreich, O. and Micciancio, D. and Safra, S. and Seifert, J. -P.},
  year = 1999,
  month = jul,
  journal = {Inf. Proc. Lett.},
  volume = {71},
  number = {2},
  pages = {55--61},
  issn = {0020-0190},
  doi = {10.1016/S0020-0190(99)00083-6},
  url = {https://www.sciencedirect.com/science/article/pii/S0020019099000836},
  urldate = {2025-10-23}
}

@article{hanggli_enhanced_2020-1,
  title = {Enhanced Noise Resilience of the Surface--{{Gottesman-Kitaev-Preskill}} Code via Designed Bias},
  author = {H{\"a}nggli, Lisa and Heinze, Margret and K{\"o}nig, Robert},
  year = 2020,
  eprint = {2004.00541},
  optprimaryclass = {quant-ph},
  archiveprefix = {arXiv},
  month = nov,
  journal = {Phys. Rev. A},
  volume = {102},
  number = {5},
  pages = {052408},
  publisher = {American Physical Society},
  doi = {10.1103/PhysRevA.102.052408},
  url = {https://link.aps.org/doi/10.1103/PhysRevA.102.052408},
  urldate = {2021-05-06}
}

@article{HarringtonRates,
  title = {Achievable rates for the {Gaussian} quantum channel},
  author = {Harrington, Jim and Preskill, John},
  year = {2001},
  eprint = {quant-ph/0105058},
  optprimaryclass = {quant-ph},
  archiveprefix = {arXiv},
  month = dec,
  journal = {Phys. Rev. A},
  volume = {64},
  number = {6},
  pages = {062301},
  publisher = {American Physical Society},
  doi = {10.1103/PhysRevA.64.062301},
  url = {https://link.aps.org/doi/10.1103/PhysRevA.64.062301}
}

@inproceedings{kurkoski_message-passing_2008,
  title = {Message-Passing Decoding of Lattices Using {{Gaussian}} Mixtures},
  booktitle = {2008 {{IEEE International Symposium}} on {{Information Theory}}},
  author = {Kurkoski, Brian and Dauwels, Justin},
  year = 2008,
  eprint = {0802.0554},
  optprimaryclass = {cs.IT},
  archiveprefix = {arXiv},
  month = jul,
  pages = {2489--2493},
  issn = {2157-8117},
  doi = {10.1109/ISIT.2008.4595439}
}

@inproceedings{kurkoski_single-gaussian_2009,
  title = {Single-{{Gaussian}} Messages and Noise Thresholds for Decoding Low-Density Lattice Codes},
  booktitle = {2009 {{IEEE International Symposium}} on {{Information Theory}}},
  author = {Kurkoski, Brian M. and Yamaguchi, Kazuhiko and Kobayashi, Kingo},
  year = 2009,
  month = jun,
  pages = {734--738},
  publisher = {IEEE},
  address = {Seoul, South Korea},
  doi = {10.1109/ISIT.2009.5205680},
  url = {http://ieeexplore.ieee.org/document/5205680/},
  urldate = {2023-10-31},
  isbn = {978-1-4244-4312-3},
  langid = {english}
}

@misc{ldlc,
  title = {Low {{density lattice codes}}},
  author = {Sommer, Naftali and Feder, Meir and Shalvi, Ofir},
  year = 2007,
  month = apr,
  eprint = {0704.1317},
  optprimaryclass = {cs, math},
  url = {http://arxiv.org/abs/0704.1317},
  urldate = {2022-07-25},
  archiveprefix = {arXiv}
}

@article{lenstra_factoring_1982,
  title = {Factoring Polynomials with Rational Coefficients},
  author = {Lenstra, A. K. and Lenstra, H. W. and Lov{\'a}sz, L.},
  year = 1982,
  month = dec,
  journal = {Math. Ann.},
  volume = {261},
  number = {4},
  pages = {515--534},
  issn = {1432-1807},
  doi = {10.1007/BF01457454},
  url = {https://doi.org/10.1007/BF01457454},
  urldate = {2025-11-05},
  langid = {english}
}

@inproceedings{li_construction_2015,
  ids = {7405587},
  title = {Construction of Structured Low Density Lattice Codes Based on Finite Fields},
  booktitle = {2015 {{IEEE Symposium}} on {{Computers}} and {{Communication}} ({{ISCC}})},
  author = {Li, Jia-Yun and Xia, Shu-Tao and Liu, Xin-Ji},
  year = 2015,
  month = jul,
  pages = {643--648},
  doi = {10.1109/ISCC.2015.7405587},
  url = {https://ieeexplore.ieee.org/document/7405587},
  urldate = {2024-08-01}
}

@article{Lin_2023,
  title = {Closest {{lattice point decoding}} for {{multimode Gottesman-Kitaev-Preskill codes}}},
  author = {Lin, Mao and Chamberland, Christopher and Noh, Kyungjoo},
  year = 2023,
  eprint = {2303.04702},
  optprimaryclass = {quant-ph},
  archiveprefix = {arXiv},
  month = dec,
  journal = {PRX Quantum},
  volume = {4},
  number = {4},
  pages = {040334},
  doi = {10.1103/PRXQuantum.4.040334},
  url = {https://link.aps.org/doi/10.1103/PRXQuantum.4.040334},
  urldate = {2023-12-03}
}

@article{liu_efficient_2019,
  title = {Efficient {{decoding}} of {{low density lattice codes}}},
  author = {Liu, Shuiyin and Hong, Yi and Viterbo, Emanuele and Marelli, Alessia and Micheloni, Rino},
  year = 2019,
  eprint = {1806.05524},
  optprimaryclass = {cs.IT},
  archiveprefix = {arXiv},
  month = aug,
  journal = {IEEE Wire. Comm. Lett.},
  volume = {8},
  number = {4},
  pages = {1195--1199},
  issn = {2162-2345},
  doi = {10.1109/LWC.2019.2911503}
}

@book{MicciancioGoldwasser:2002,
  title = {Complexity of lattice problems: A cryptographic perspective},
  author = {Micciancio, Daniele and Goldwasser, Shafi},
  year = {2002},
  series = {The Springer International Series in Engineering and Computer Science},
  volume = {671},
  publisher = {Springer},
  address = {Boston, MA},
  isbn = {978-0-7923-7688-0},
  doi = {10.1007/978-1-4615-0897-7},
  url = {https://doi.org/10.1007/978-1-4615-0897-7}
}

@article{Noh_2019,
  title = {Fault-Tolerant Bosonic Quantum Error Correction with the Surface--{{Gottesman-Kitaev-Preskill}} Code},
  author = {Noh, Kyungjoo and Chamberland, Christopher},
  year = 2020,
  eprint = {1908.03579},
  optprimaryclass = {quant-ph},
  archiveprefix = {arXiv},
  month = jan,
  journal = {Phys. Rev. A},
  volume = {101},
  number = {1},
  pages = {012316},
  publisher = {American Physical Society},
  doi = {10.1103/PhysRevA.101.012316},
  url = {https://link.aps.org/doi/10.1103/PhysRevA.101.012316},
  urldate = {2020-04-16}
}

@article{panteleev_degenerate_2021,
  title = {Degenerate {{quantum LDPC codes with good finite length performance}}},
  author = {Panteleev, Pavel and Kalachev, Gleb},
  year = 2021,
  eprint = {1904.02703},
  optprimaryclass = {quant-ph},
  archiveprefix = {arXiv},
  month = nov,
  journal = {Quantum},
  volume = {5},
  pages = {585},
  publisher = {Verein zur F\"orderung des Open Access Publizierens in den Quantenwissenschaften},
  doi = {10.22331/q-2021-11-22-585},
  url = {https://quantum-journal.org/papers/q-2021-11-22-585/},
  urldate = {2023-02-15},
  langid = {british}
}

@book{papoulis_probability_2002,
  title = {Probability, random variables, and stochastic processes},
  author = {Papoulis, Athanasios and Pillai, S. Unnikrishna},
  year = 2002,
  series = {{{McGraw-Hill}} Series in Electrical and Computer Engineering},
  edition = {4th ed},
  publisher = {McGraw-Hill},
  address = {Boston},
  url = {http://catdir.loc.gov/catdir/enhancements/fy1011/2001044139-t.html},
  urldate = {2025-12-08},
  isbn = {978-0-07-366011-0},
  langid = {english}
}

@article{PhysRevA.101.053840,
  ids = {weigand_realizing_2019},
  title = {Realizing Modular Quadrature Measurements via a Tunable Photon-Pressure Coupling in Circuit {{QED}}},
  author = {Weigand, Daniel J. and Terhal, Barbara M.},
  year = 2020,
  month = may,
  journal = {Phys. Rev. A},
  volume = {101},
  number = {5},
  eprint = {1909.10075},
  optprimaryclass = {quant-ph},
  pages = {053840},
  doi = {10.1103/PhysRevA.101.053840},
  url = {https://link.aps.org/doi/10.1103/PhysRevA.101.053840},
  urldate = {2025-04-30},
  archiveprefix = {arXiv}
}

@article{PhysRevLett.128.170503,
  title = {Protocol for generating optical {Gottesman--Kitaev--Preskill} states with cavity {QED}},
  author = {Hastrup, Jacob and Andersen, Ulrik L.},
  year = {2022},
  eprint = {2104.07981},
  optprimaryclass = {quant-ph},
  archiveprefix = {arXiv},
  month = apr,
  journal = {Phys. Rev. Lett.},
  volume = {128},
  number = {17},
  pages = {170503},
  publisher = {American Physical Society},
  doi = {10.1103/PhysRevLett.128.170503},
  url = {https://link.aps.org/doi/10.1103/PhysRevLett.128.170503}
}

@article{PhysRevX.8.021054,
  title = {High-Threshold Fault-Tolerant Quantum Computation with Analog Quantum Error Correction},
  author = {Fukui, Kosuke and Tomita, Akihisa and Okamoto, Atsushi and Fujii, Keisuke},
  year = 2018,
  eprint = {1712.00294},
  optprimaryclass = {quant-ph},
  archiveprefix = {arXiv},
  month = may,
  journal = {Phys. Rev. X},
  volume = {8},
  number = {2},
  pages = {021054},
  publisher = {American Physical Society},
  doi = {10.1103/PhysRevX.8.021054},
  url = {https://link.aps.org/doi/10.1103/PhysRevX.8.021054}
}

@article{poltyrev_coding_1994,
  title = {On Coding without Restrictions for the {{AWGN}} Channel},
  author = {Poltyrev, G.},
  year = 1994,
  month = mar,
  journal = {IEEE Trans. Inf. Th.},
  volume = {40},
  number = {2},
  pages = {409--417},
  issn = {1557-9654},
  doi = {10.1109/18.312163},
  url = {https://ieeexplore.ieee.org/document/312163/},
  urldate = {2025-05-20}
}

@article{PRXQuantum.3.010315,
  title = {Low-{{overhead fault-tolerant quantum error correction}} with the {{surface-GKP code}}},
  author = {Noh, Kyungjoo and Chamberland, Christopher and Brand{\~a}o, Fernando G.S.L.},
  year = 2022,
  month = jan,
  journal = {PRX Quantum},
  volume = {3},
  number = {1},
  eprint = {2103.06994},
  pages = {010315},
  publisher = {American Physical Society},
  doi = {10.1103/PRXQuantum.3.010315},
  url = {https://link.aps.org/doi/10.1103/PRXQuantum.3.010315},
  urldate = {2022-02-09},
  archiveprefix = {arXiv}
}

@article{PRXQuantum.4.020342,
  ids = {xu_qubit-oscillator_2022},
  title = {Qubit-{{oscillator concatenated codes}}: {{Decoding formalism}} and {{code comparison}}},
  shorttitle = {Qubit-{{Oscillator Concatenated Codes}}},
  author = {Xu, Yijia and Wang, Yixu and Kuo, En-Jui and Albert, Victor V.},
  year = 2023,
  eprint = {2209.04573},
  optprimaryclass = {quant-ph},
  archiveprefix = {arXiv},
  month = jun,
  journal = {PRX Quantum},
  volume = {4},
  number = {2},
  pages = {020342},
  publisher = {American Physical Society},
  doi = {10.1103/PRXQuantum.4.020342},
  url = {https://link.aps.org/doi/10.1103/PRXQuantum.4.020342},
  urldate = {2024-10-28}
}

@article{PRXQuantum.5.020349,
  title = {Analog {{information decoding}} of {{bosonic quantum low-density parity-check codes}}},
  shorttitle = {Bosonic {{QLPDC}}},
  author = {Berent, Lucas and Hillmann, Timo and Eisert, Jens and Wille, Robert and Roffe, Joschka},
  year = 2024,
  month = may,
  journal = {PRX Quantum},
  volume = {5},
  number = {2},
  eprint = {2311.01328},
  optprimaryclass = {quant-ph},
  pages = {020349},
  publisher = {American Physical Society},
  doi = {10.1103/PRXQuantum.5.020349},
  url = {https://link.aps.org/doi/10.1103/PRXQuantum.5.020349},
  urldate = {2024-05-31},
  archiveprefix = {arXiv}
}

@article{QECBasic,
  title = {Quantum error correction: An introductory guide},
  author = {Roffe, Joschka},
  year = {2019},
  eprint = {1907.11157},
  optprimaryclass = {quant-ph},
  archiveprefix = {arXiv},
  journal = {Contemp. Phys.},
  volume = {60},
  number = {3},
  pages = {226--245},
  doi = {10.1080/00107514.2019.1667078},
  url = {https://doi.org/10.1080/00107514.2019.1667078}
}

@article{QLDPCGKP,
  title = {Finite {{Rate QLDPC-GKP coding scheme}} that {{surpasses}} the {{CSS Hamming bound}}},
  author = {Raveendran, Nithin and Rengaswamy, Narayanan and Rozp{\k e}dek, Filip and Raina, Ankur and Jiang, Liang and Vasi{\'c}, Bane},
  year = 2022,
  eprint = {2111.07029},
  optprimaryclass = {quant-ph},
  archiveprefix = {arXiv},
  month = jul,
  journal = {Quantum},
  volume = {6},
  pages = {767},
  publisher = {Verein zur F\"orderung des Open Access Publizierens in den Quantenwissenschaften},
  doi = {10.22331/q-2022-07-20-767},
  url = {https://quantum-journal.org/papers/q-2022-07-20-767/},
  urldate = {2023-11-02},
  langid = {british}
}

@article{ReviewGKP,
  ids = {brady_advances_2023},
  title = {Advances in Bosonic Quantum Error Correction with {{Gottesman}}--{{Kitaev}}--{{Preskill codes}}: {{Theory}}, Engineering and Applications},
  shorttitle = {Advances in Bosonic Quantum Error Correction with {{Gottesman}}--{{Kitaev}}--{{Preskill Codes}}},
  author = {Brady, Anthony J. and Eickbusch, Alec and Singh, Shraddha and Wu, Jing and Zhuang, Quntao},
  year = 2024,
  month = jan,
  journal = {Progr. Quant. Elect.},
  volume = {93},
  eprint = {2308.02913},
  optprimaryclass = {quant-ph},
  pages = {100496},
  issn = {0079-6727},
  doi = {10.1016/j.pquantelec.2023.100496},
  url = {https://www.sciencedirect.com/science/article/pii/S0079672723000459},
  urldate = {2025-04-30},
  archiveprefix = {arXiv}
}

@article{RevModPhys.87.307,
  title = {Quantum Error Correction for Quantum Memories},
  author = {Terhal, Barbara M.},
  year = 2015,
  eprint = {1302.3428},
  optprimaryclass = {quant-ph},
  archiveprefix = {arXiv},
  month = apr,
  journal = {Rev. Mod. Phys.},
  volume = {87},
  number = {2},
  pages = {307--346},
  publisher = {American Physical Society},
  doi = {10.1103/RevModPhys.87.307},
  url = {https://link.aps.org/doi/10.1103/RevModPhys.87.307},
  urldate = {2020-07-14}
}

@article{Roads,
  title = {Roads towards Fault-Tolerant Universal Quantum Computation},
  author = {Campbell, Earl T. and Terhal, Barbara M. and Vuillot, Christophe},
  year = 2017,
  eprint = {1612.07330},
  optprimaryclass = {quant-ph},
  archiveprefix = {arXiv},
  month = sep,
  journal = {Nature},
  volume = {549},
  number = {7671},
  pages = {172--179},
  publisher = {Nature Publishing Group},
  issn = {1476-4687},
  doi = {10.1038/nature23460},
  url = {https://www.nature.com/articles/nature23460},
  urldate = {2021-02-15},
  copyright = {2017 Macmillan Publishers Limited, part of Springer Nature. All rights reserved.},
  langid = {english}
}

@article{shannon_mathematical_1948,
  title = {A Mathematical Theory of Communication},
  author = {Shannon, C. E.},
  year = 1948,
  month = jul,
  journal = {The Bell Syst. Tech. J.},
  volume = {27},
  number = {3},
  pages = {379--423},
  issn = {0005-8580},
  doi = {10.1002/j.1538-7305.1948.tb01338.x},
  url = {https://ieeexplore.ieee.org/document/6773024},
  urldate = {2025-10-24}
}

@article{Spencer2026quditlowdensity,
	doi = {10.22331/q-2026-03-13-2023},
	url = {https://doi.org/10.22331/q-2026-03-13-2023},
	title = {Qudit low-density parity-check codes},
	author = {Spencer, Daniel J. and Tanggara, Andrew and Haug, Tobias and Khu, Derek and Bharti, Kishor},
	journal = {{Quantum}},
	issn = {2521-327X},
	publisher = {{Verein zur F{\"{o}}rderung des Open Access Publizierens in den Quantenwissenschaften}},
	volume = {10},
	pages = {2023},
	month = mar,
	year = {2026},
  eprint = {2510.06495},
  optprimaryclass = {quant-ph},
  archiveprefix = {arXiv},
}

@article{toricGKP,
  title = {Quantum Error Correction with the Toric {{Gottesman-Kitaev-Preskill}} Code},
  author = {Vuillot, Christophe and Asasi, Hamed and Wang, Yang and Pryadko, Leonid P. and Terhal, Barbara M.},
  year = 2019,
  eprint = {1810.00047},
  optprimaryclass = {quant-ph},
  archiveprefix = {arXiv},
  month = mar,
  journal = {Phys. Rev. A},
  volume = {99},
  number = {3},
  pages = {032344},
  publisher = {American Physical Society},
  doi = {10.1103/PhysRevA.99.032344},
  url = {https://link.aps.org/doi/10.1103/PhysRevA.99.032344},
  urldate = {2020-04-16}
}

@article{wang_efficient_2023,
  title = {Efficient {{decoder design}} for {{low-density lattice codes from}} the {{lattice viewpoint}}},
  author = {Wang, Xuebo and Mow, Wai Ho},
  year = 2023,
  journal = {IEEE Open J. Commun. Soc.},
  volume = {4},
  pages = {1839--1854},
  issn = {2644-125X},
  doi = {10.1109/OJCOMS.2023.3305480},
  url = {https://ieeexplore.ieee.org/document/10217184/},
  urldate = {2023-10-06},
  langid = {english}
}

@misc{webster_explicit_2025,
  title = {Explicit Construction of Low-Overhead Gadgets for Gates on Quantum {{LDPC}} Codes},
  author = {Webster, Paul and Smith, Samuel C. and Cohen, Lawrence Z.},
  year = 2025,
  month = nov,
  number = {arXiv:2511.15989},
  eprint = {2511.15989},
  optprimaryclass = {quant-ph},
  publisher = {arXiv},
  doi = {10.48550/arXiv.2511.15989},
  url = {http://arxiv.org/abs/2511.15989},
  archiveprefix = {arXiv}
}

@article{wen_efficient_2016,
  title = {An {{efficient algorithm}} for {{optimally solving}} a {{shortest vector problem}} in {{compute-and-forward design}}},
  author = {Wen, Jinming and Zhou, Baojian and Mow, Wai Ho and Chang, Xiao-Wen},
  year = 2016,
  eprint = {1410.4278},
  optprimaryclass = {cs.IT},
  archiveprefix = {arXiv},
  month = oct,
  journal = {IEEE Trans. Wire. Comm.},
  volume = {15},
  number = {10},
  pages = {6541--6555},
  issn = {1558-2248},
  doi = {10.1109/TWC.2016.2585493}
}

@book{weyl1950theory,
  title = {The theory of groups and quantum mechanics},
  author = {Weyl, Hermann},
  year = {1950},
  series = {Dover Books on Mathematics},
  publisher = {Dover Publications},
  address = {New York},
  isbn = {978-0-486-60269-1}
}

\end{document}